\documentclass[final]{JFM-FLM_Au}

\usepackage{multirow}%
\usepackage{gensymb}

\newcommand*\dist{\includegraphics[width=3.5mm]{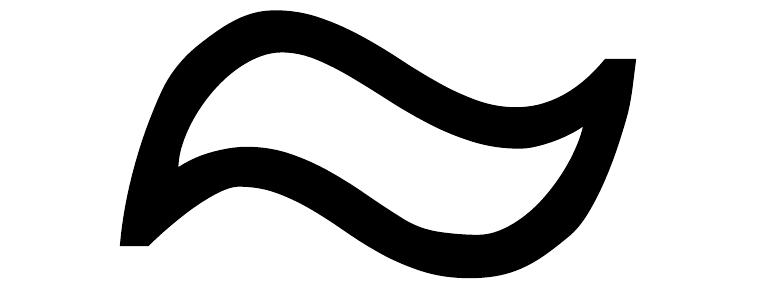}}
\newcommand{\uniform}[2]{U\left({#1},{#2}\right)}
\newcommand{\triangledist}[2]{T\left({#1},{#2}\right)}
\newcommand{\triangledistmode}[3]{T\left({#1},{#2},\text{mode=}{#3}\right)}

\newcommand{\vvec}[1]{\vec{\vec{#1}}}
\DeclareMathOperator{\acos}{\cos^{-1}}
\DeclareMathOperator{\asin}{\sin^{-1}}

\lefttitle{M.M. Flapper, G. Piumini, S.G. Huisman, R. Verzicco, and D. Lohse}
\righttitle{Journal of Fluid Mechanics}

\title{Settling dynamics of an oloid: experiments and simulations}

\author{Mees M. Flapper\aff{1}, Giulia Piumini\aff{1}, Roberto Verzicco\aff{1,2}, Sander G. Huisman\aff{1}, and Detlef Lohse\aff{1,3}}

\affiliation{\aff{1}Physics of Fluids Department and Max Planck Center for Complex Fluid Dynamics, Faculty of Science \& Technology, University of Twente, Drienerlolaan 5, 7500NB, Enschede, The Netherlands
\aff{2}Dipartimento di Ingegneria Industriale, University of Roma ‘Tor Vergata’, Via del Politecnico 1, 00133, Roma, Italy
\aff{3}Max Planck Institute for Dynamics and Self-Organization, Am Fassberg 17, 37077, Göttingen, Germany}

\corresau{D. Lohse, \email{d.lohse@utwente.nl}}

\begin{document}
\maketitle

\begin{abstract}
This study presents a combined experimental and computational investigation of an oloid-shaped particle settling in a quiescent fluid. The oloid, a unique convex shape with anisotropic geometry, provides a distinctive model for exploring how a particle's shape and orientation affect its settling dynamics. The settling oloids are tracked experimentally for Galileo numbers $48 \leq \mbox{\text{Ga}} \leq 5.4 \cdot 10^3$, using two particle sizes ($D_{\text{eq}} =$ 21.6 mm, and $D_{\text{eq}} =$ 10.8 mm). The density ratio between the particle and fluid $\Gamma = \frac{\rho_p}{\rho_f}$ ranges from $1.11 \leq \Gamma \leq 1.30$ in the experiments. Computationally, the Galileo numbers $10 \leq \mbox{\text{Ga}} \leq 100$ are simulated, with $\Gamma = 2$. The experimental findings and numerical results are in good agreement, and give a consistent idea of the oloid settling dynamics. Our results indicate two distinct falling modes for the oloid, separated by Galileo number. The stable mode is characterised by a preferential orientation, with a rotation around the vertical axis, whereas the tumbling mode has randomly distributed orientation and rotation statistics. We characterise the falling velocity, orientation, and rotation dynamics of the oloids over a range of Galileo numbers. Additionally, the influence of the initial orientation is revealed to determine the rotation dynamics at low Galileo numbers.
\end{abstract}


\section{Introduction}
Settling experiments in still fluids provide insight into the motion of particles under the influence of gravity, and have natural and industrial applications, including recycling processes based on density separation \citep{Bauer2018}, sedimentary rocks and structures in sedimentology \citep{nichols2009sedimentology}, volcanoclastic sediments \citep{Rios2023}, pollutants in the air \citep{Hinds2022}, the spreading of seeds \citep{lentink2009} and marine snow \citep{alldredge1988}. 
The settling of heavy spheres has been extensively studied, with numerical and experimental investigations showing the existence of multiple falling regimes \citep{Jenny2004,Veldhuis2007,Horowitz2010,Uhlmann2014}. These studies identified four distinct regimes: steady vertical, steady oblique, oblique oscillating, and chaotic. It is the Galileo number which determines the falling mode of the sphere, with a slight dependence on the density ratio between particle and fluid \citep{Veldhuis2007}.

A comparison with buoyant spheres shows similarities, with spheres rising in a rectilinear trajectory (straight or oblique), or zigzagging periodically. The mass ratio and the Reynolds number determine the rising dynamics in this case \citep{Horowitz2010}.
However, spherical, isotropic particles are the exception; in general, particles deviate from this, resulting in a wide range of different shapes, which can be anisotropic. The latter have recently been studied, thus adding more complexity owing to the role of shape and orientation in settling dynamics \citep{Guazzelli2011,Voth2017,mathai2018}. 
Also here, understanding the settling of such anisotropic particles is crucial for predicting the dynamics of particles in natural and industrial processes, such as sediment deposition \citep{Zouaoui}, producing particulate systems such as food, batteries and pharmaceuticals \citep{Ulusoy2023}, or particle aggregation in the ocean, which affects plankton dynamics and other ecological processes in marine environments \citep{Burd}.

The settling velocity of anisotropic particles is influenced by factors like shape, orientation, density, aspect ratio, and mass distribution \citep{Roy2019}. Furthermore, unlike spherical particles, anisotropic particles experience variable drag forces, depending on their orientation \citep{Voth2017}. Fibres or long cylinders follow a rectilinear or fluttering (zigzagging) trajectory, with periodic oscillations \citep{Toupoint2019}. Similar results have been found for disks, which also display multiple falling modes. Both fluttering and tumbling regimes have been found, separated by a chaotic regime \citep{Auguste2013}. For these disks, the falling mode depends on the aspect ratio, the particle-to-fluid density ratio, the Archimedes number and the dimensionless inertia ratio. In the previously described cases, the particle wake plays an important role, the onset of the wake instability being closely related to that of the path instability \citep{Ern2012}. Besides disks and fibres, numerical simulations of settling ellipsoids show similar results: the settling ellipsoids display a steady, oscillatory, and chaotic falling mode \citep{Fonseca2005,Moriche2021}. The tumbling falling mode is notably absent for the ellipsoids when compared to disks.

Expanding on these shapes, more complex particles exhibit yet another range of settling dynamics. \cite{Chan2021} found that heavy curved sheets zigzag as they settle in a quiescent liquid. \cite{Candelier2016} showed that a dumbbell (two spheres connected by a rod) settles horizontally as long as the two spheres are identical. However, when making one sphere slightly larger (keeping the spheres' mass density the same), the dumbbell settles at an equilibrium angle determined by the difference in sphere size and Reynolds number. For larger size differences between the spheres, the dumbbell settles vertically, with its orientation aligned to the direction of gravity.

Numerical simulations by \cite{Piumini2024} show that simple chiral particles rotate as they settle, indicating a translation-rotation coupling for these particles. At higher Reynolds number, the turbulent fluctuations dominate over the particle forcing, resulting in more isotropic dynamics.

\cite{Huseby2025} recently investigated helical ribbons, which also exhibit a strong translation-rotation coupling, and create a quasiperiodic settling trajectory. Computational results show that the ribbon trajectory is sensitive to its initial orientation, and is affected by the helical ribbon length, providing a diverse range of trajectories.

Such translation-rotation coupling was not observed for an isotropic helicoid, despite predictions that the particle should rotate as it settles \citep{Collins2021}, even though the particle is rotation symmetric. For non-chiral particles, \cite{Miara2024} showed that a rigid U-shaped disk can display a helical sedimentation path based on its initial orientation, despite the particle geometry being non-chiral. The settling path of the U-shaped disk is shown to depend on the particle's initial orientation: many different settling paths and settling dynamics are observed when varying the initial orientation.

These recent findings show a diverse range of particle dynamics, with particles showing distinct falling regimes by either changing the particle geometry, Reynolds number, or initial orientation. This provides an idea of the large parameter space and the rich dynamics in particle settling, and much of the parameter space is yet to be explored.

This work investigates the settling dynamics of an oloid-shaped particle in a quiescent fluid. An oloid is a three-dimensional curved shape discovered by Paul Schatz in 1929 \citep{oloid1}. It is a ruled surface formed as the convex hull of a frame made by placing two identical discs in perpendicular planes \citep{oloid}.
\begin{figure}
    \centering
    \includegraphics[width=\textwidth]{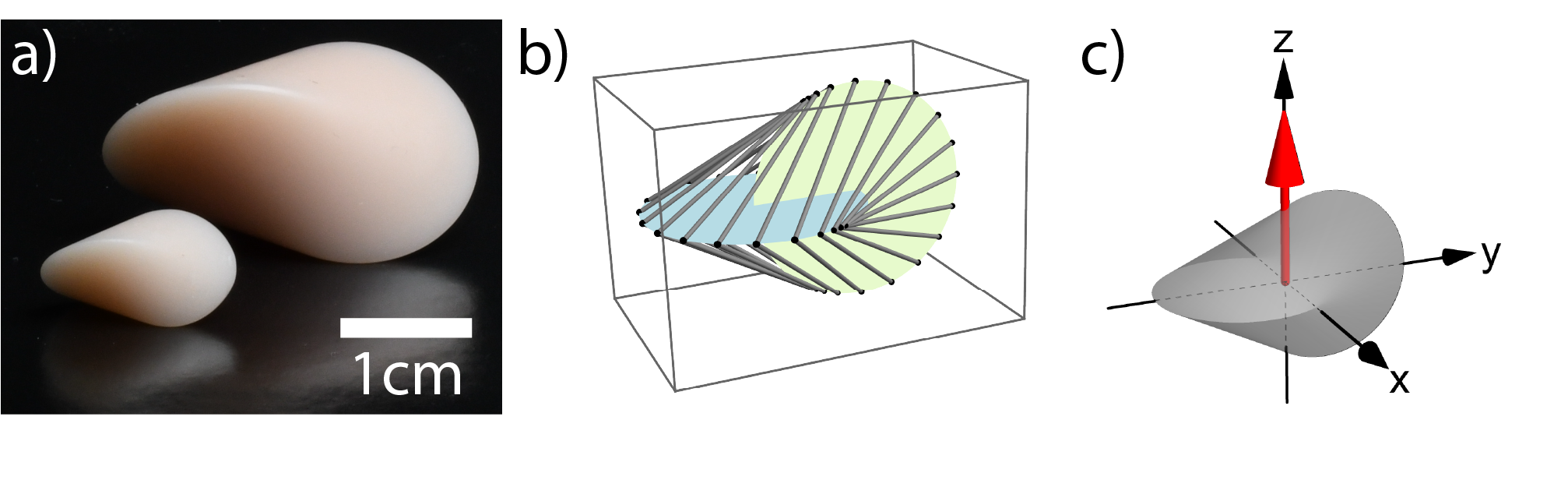}
    \caption{Photo of the 3D-printed oloid particles used in experiments (a), and a 3D visualisation of the oloid geometry (b), showing perpendicular discs intersecting through their centres, with lines illustrating the convex hull. The oloid in its reference orientation (c), where the red arrow indicates the chosen pointing vector used in our analysis.}
    \label{fig:Oloid_photo_model}
\end{figure}
Figure \ref{fig:Oloid_photo_model}a) shows the 3D-printed oloids used in the experiments. Figure \ref{fig:Oloid_photo_model}b) shows a visualisation of the oloid geometry, displaying the discs used to define and generate the shape. The lines connecting the discs indicate the convex hull of the connected discs. Finally, figure \ref{fig:Oloid_photo_model}c) displays the oloid in its reference orientation, with the red arrow indicating the pointing vector chosen for our analysis. When placed on an inclined flat surface, the oloid geometry leads to a characteristic rolling motion, where every point on its surface touches the wall at some stage of its motion \citep{oloid}. Despite the geometrical simplicity, the oloid displays complex dynamics due to its interplay of symmetry and directional dependence. Its unique geometry has practical applications as well, making it useful for tasks like water treatment, propeller design, and as an efficient stirrer \citep{ol2}. Additionally, the oloid shape has recently been used to create a magnetically driven robot, to be used inside the human body \citep{Greenidge2025}. This oloid-shaped robot was able to identify lesions, and made detailed three-dimensional scans of subsurface tissue, showing the applicability of the oloid shape in robotics and healthcare.

In this paper we look into the oloid's settling dynamics in a quiescent fluid, both computationally and experimentally. The Galileo number is varied for both experiments and simulations, to explore the influence of this parameter, and probe whether multiple falling regimes exist. The orientation and position of the oloid is carefully tracked, allowing the close investigation of the settling trajectory and corresponding orientation dynamics. These combined results reveal whether any translation-rotation coupling exists for these particles. \\
This paper is organised as follows: the experimental setup is described in section \ref{sec:Experimental setup}, whereas the numerical setup is discussed in section \ref{sec:Numerical setup}. The combined results of the experimental and numerical investigations are shown in section \ref{sec:Results}. We close with conclusions and an outlook in section \ref{sec:Outlook and conclusions}.

\section{Experimental setup} \label{sec:Experimental setup}
\begin{figure}
    \centering
    \includegraphics[width=0.4\linewidth]{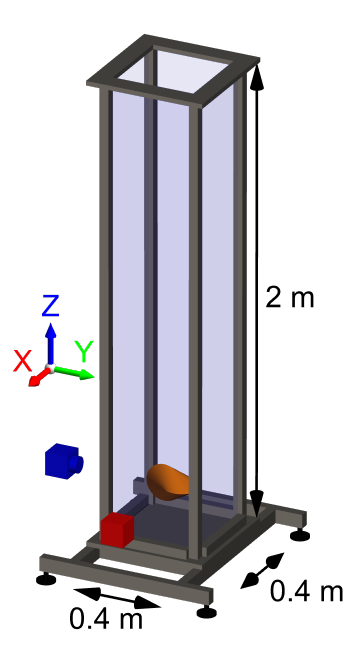}
    \caption{3D visualisation of the settling tank used in experiments, along with the cameras shown in red and blue, and an oloid (not to scale) in orange. The dimensions of the settling tank are $h =$ 200 cm, $l = w =$ 40 cm, where $h$, $l$, and $w$ denote the height, length, and width of the tank (without supports), respectively.}
    \label{fig:setup}
\end{figure}

The experimental investigations of the settling oloid are performed in a large rectangular settling tank, illustrated in figure \ref{fig:setup}, where the blue and red cuboids represent high-speed cameras. 
The height of the setup is $h =$ 200 cm, the width and height are $l = w =$ 40 cm. The tracked oloids are 3D-printed in two formats: the larger oloid weighs 7.1 g and its size, characterised by the radius of the generating disk, measures 12.0 mm. This gives an equivalent diameter $D_{\text{eq}} =$ 21.6 mm, which represents the diameter of a sphere of equal volume. The smaller oloid weighs 0.85 g and has a characteristic radius 6.0 mm, or $D_{\text{eq}} =$ 10.8 mm. Both oloids are made of FormLabs Model resin, with a density of $\rho =$ $1.30 \cdot 10^3$ kg/m$^3$. The oloids used for the experiments are shown in figure \ref{fig:Oloid_photo_model}a).

Two Photron Mini AX-200 high-speed cameras are set up orthogonally to track the settling oloid at a frame rate of 50 fps up to 250 fps, depending on the Galileo number. The calibrated measurement volume has dimensions 33 cm $\times$ 33 cm $\times$ 42 cm. The cameras record at 1024 px $\times$ 1024 px, with the resolution for the two cameras around 325 $\mu$m/px and 275$\mu$m/px respectively.

\begin{figure}
    \centering
    \includegraphics[width=0.9\textwidth]{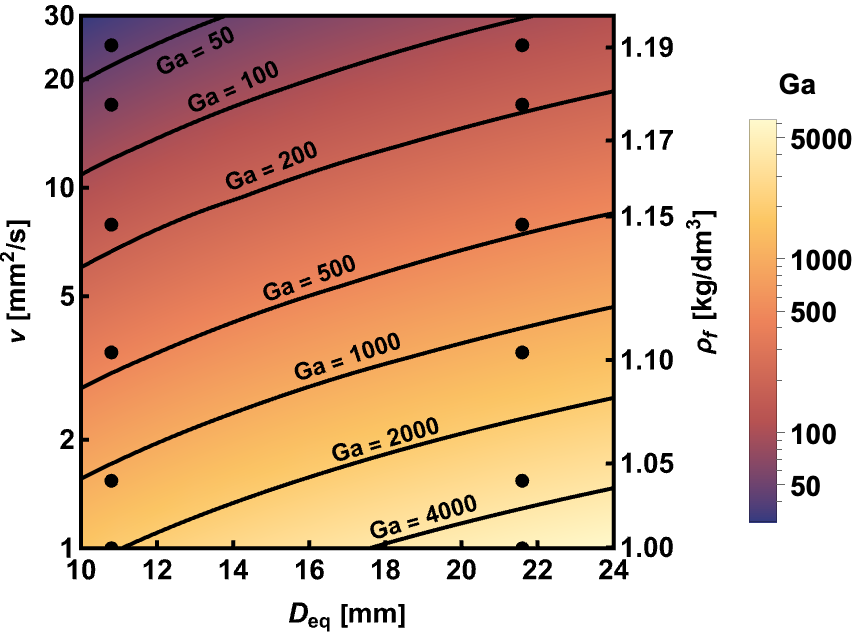}
    \caption{Plot of the parameter space of the oloid in water-glycerol mixture, based on documented properties of glycerine solutions at 20 \degree C \citep{glycerine1963}. The used oloids have an equivalent diameter of $D_{\text{eq}} =$ 10.8 mm, and $D_{\text{eq}} =$ 21.6 mm. The black dots show the parameter values used in experiments.}
    \label{fig:Ga_plot}
\end{figure}

The particle settles in different water-glycerol mixtures, which gives a range of Galileo (and particle Reynolds) numbers, defined as
\begin{equation}
    \text{Ga} = \frac{D_{\text{eq}} \sqrt{g D_{\text{eq}} \left(\frac{\rho_p}{\rho_f} - 1 \right)}}{\nu}.
\end{equation}
Here $g$ is the gravitational acceleration, $\nu$ is the fluid kinematic viscosity, and $\rho_p$ and $\rho_f$ are the density of the particle and fluid respectively. This definition of the Galileo number (including the density difference) is equivalent to the square root of the conventional Archimedes number. We vary the Galileo number through the kinematic viscosity $\nu$ and the density $\rho_f$, by modifying the amount of glycerol in the water-glycerol mixture. The accessible parameter space for the oloids in water-glycerol mixtures is shown in figure \ref{fig:Ga_plot}, with the black points indicating the parameters at which we performed experiments.
The used water-glycerol mixtures and resulting Galileo numbers for the experiments are shown in table \ref{tab:Ga_numbers},
indicating the wide range of experimentally investigated Galileo numbers. The viscosity of the mixtures with $40\%$ glycerol and higher is determined using an Anton Paar MCR 502 rheometer, whereas the viscosity of lower glycerol fractions is found using the values reported for glycerol properties \citep{glycerine1963}. The fluid density is measured by taking a sample of each water-glycerol mixture directly after completing the experiment.
\begin{table}
\centering
\begin{tabular}{cccccc}
\hline
& \begin{tabular} [c]{@{}c@{}}Glycerol fraction\\ by weight [\%]\end{tabular} & $\nu$ [$\text{mm}^2/\text{s}$] & $\Gamma = \frac{\rho_p}{\rho_f}$ & \begin{tabular}[c]{@{}c@{}}Ga\\ Small Oloid\end{tabular} & \begin{tabular}[c]{@{}c@{}}Ga\\ Large oloid\end{tabular} \\ \hline
\multirow{6}{4em}{Experiments} & $0$  & $1.0$  &  $1.30$ & $1900$ & $5400$ \\ 
& $17$ & $1.5$ & $1.27$ & $1200$ & $3500$ \\ 
& $40$ & $3.5$ & $1.19$ & $440$ & $1300$ \\ 
& $56$ & $7.9$ & $1.15$ & $170$ & $490$ \\ 
& $67$ & $17$ & $1.13$ & $75$ & $210$ \\ 
& $74$ & $25$ & $1.11$ & $48$ & $130$ \\ \hline
&   & $\nu$  & $\Gamma = \frac{\rho_p}{\rho_f}$  & \multicolumn{2}{c}{Ga} \\ \hline
\multirow{3}{4em}{Simulations} &   & $0.1$  &  $2$ & \multicolumn{2}{c}{10} \\
&   & $0.02$  &  $2$  &  \multicolumn{2}{c}{50} \\
&   & $0.01$  &  $2$ &  \multicolumn{2}{c}{100} \\ \hline
\end{tabular}
\caption{Parameters for the water-glycerol mixtures used in the experiments, with the associated Galileo numbers for both the small and large oloid. The viscosity of the fluids with a glycerol fraction of $40\%$ and higher was measured using a rheometer (Anton Paar MCR 502), the viscosity for the mixtures with smaller glycerol fractions is based on documented glycerine properties \citep{glycerine1963}.}
\label{tab:Ga_numbers}
\end{table}
For each Galileo number, 10 measurements were performed. The oloid was released by placing it on a small platform, submerging the platform and oloid, and letting the oloid fall from the submerged platform. The oloid was released near the water surface, but only recorded near the bottom of the settling tank, ensuring that any transients have died out, and that the particle is at the terminal falling velocity. By releasing the oloid in this manner, the oloid's initial orientation cannot be accurately controlled. Therefore we did not record it; the effects of the oloid starting orientation are only studied in simulations.

\begin{figure}
    \centering
    \includegraphics[width=0.9\textwidth]{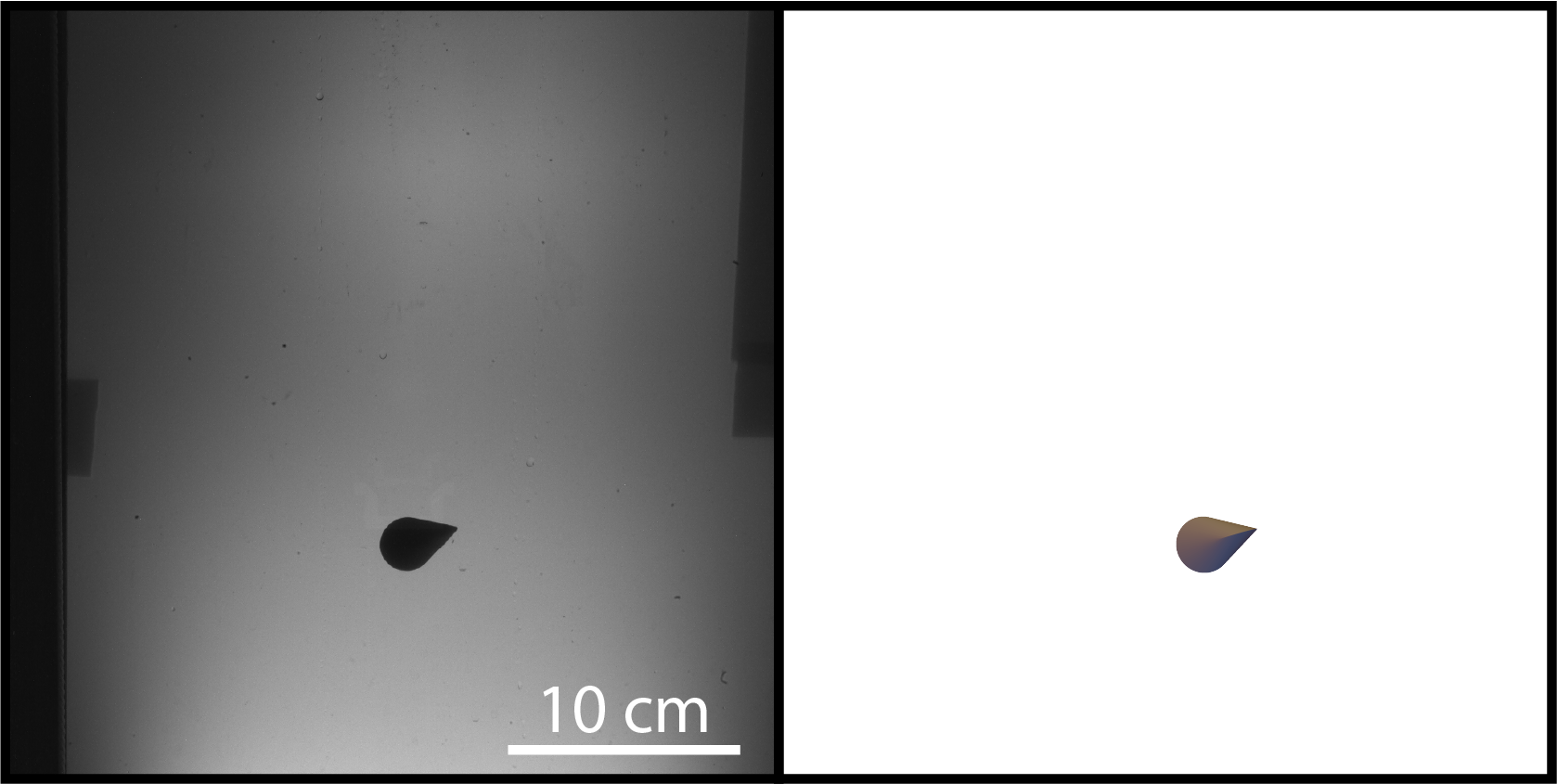}
    \caption{A raw image of an oloid captured by one of the high-speed cameras (left panel), and a reconstruction of the oloid using our orientation tracking algorithm described in \cite{Flapper2025} (right panel). An animation of the camera recording and the reconstruction is shown in the supplementary material.}
    \label{fig:oloid_reconstruction}
\end{figure}

The location and orientation of the settling oloids were tracked using the high-speed recordings as described in \cite{Flapper2025}. The location of the particles was found by centroid matching, whereas the orientation was found by matching the recorded particles to the known oloid geometry. In short, the recorded projections of the oloid were matched to projections of a synthetic oloid. The difference between the experimental and synthetic projections was minimised (using a Nelder--Mead algorithm to vary the synthetic particle's orientation). The resulting orientation of the synthetic oloid then gives the orientation of the experimentally tracked oloid. An example of the detected and reconstructed oloid orientation is shown in figure \ref{fig:oloid_reconstruction}, displaying the raw camera image on the left, and the reconstructed oloid on the right. An animation of a reconstructed oloid alongside the raw camera recording is shown in the supplementary material.
Using this orientation tracking method, the rotation and orientation dynamics of the oloid can be studied over time.

\section{Numerical setup}
\label{sec:Numerical setup}
\noindent The governing relations for the fluid are the incompressible Navier--Stokes and continuity equations, which in nondimensional form read
\begin{equation}
\frac{\partial\bold{u}}{\partial t} + (\bold{u}\cdot \boldsymbol{\nabla})\bold{u} =
-\boldsymbol{\nabla} p  +\frac{1}{\text{Re}}\boldsymbol{\nabla}^2 \bold{u} + \mathbf{f},\ \ \ \ \
\boldsymbol{\nabla} \cdot \bold{u} = 0.
\label{eq:ch3_NS}
\end{equation}
Here, $\bold{u}$ and $p$  are fluid velocity vector and kinematic pressure, respectively, while \Rey$=D_{\text{eq}}U_g/\nu$ is the Reynolds number
based on the diameter of a sphere of equal volume $D_{\text{eq}}$ and the velocity $U_g=({ D_{\text{eq}} |\mathbf{g}|(1-{1}/{\Gamma})})^{1/2}$. The volume force $\mathbf{f}$ at the right hand side of equations \ref{eq:ch3_NS} is the immersed boundary method contribution that accounts for the presence of solid particles, allowing for the two-way coupling between dispersed and carrier phase, as in previous research by \cite{Piumini2024}

The dynamics of the oloid are obtained by solving the Newton--Euler equations, following \cite{BREUGEM20124469} and relying on Newton's third law of motion, yielding
  \begin{align}
    \begin{split}
        \displaystyle \frac{d\mathbf{v}_c}{dt}= &\frac{1}{\Gamma}\frac{6}{\pi} \Bigg( -\sum_{i=1}^{N_l} \mathbf{f}_i\Delta V_i+\frac{d}{dt} \int_{V_p}\mathbf{v}dV\Bigg)-\frac{\widehat{{\bf k}}}{\textit{Fr}}, \\
    \displaystyle \ \mathbf{I}_p\frac{d\boldsymbol{\omega}_c}{dt}+\boldsymbol{\omega}_c\times(\mathbf{I}_p\boldsymbol{\omega}_c)= &\frac{1}{\Gamma}\frac{6}{\pi}\Bigg(- \sum_{i=1}^{N_l} \mathbf{r}_i^n\times \mathbf{f}_i\Delta V_i+\frac{d}{dt}\int_{V_p}\mathbf{r}\times \mathbf{v}dV\Bigg).
\label{eq:ch3_newton}
\end{split}
\end{align}
Here $\mathbf{v}_c$ and $\boldsymbol{\omega}_c$ are the particle centre of mass linear and angular velocities. $\mathbf{I}_p$ is the particle inertial tensor that, in the principal inertial reference frame,
has only the diagonal
components.  $\widehat{{\bf k}}$ is the vertical unit vector (anti-parallel to gravity), and $\textit{Fr}= {U^2}/({ D_{\text{eq}} |\mathbf{g}|(1-{1}/{\Gamma})})$ is the Froude number.
The particle surface is triangulated and each triangle is tagged by a Lagrangian marker at its centroid; thus in the above equations \eqref{eq:ch3_newton} the index $i$ indicates the Lagrangian point over
the surface and $\Delta V_i$ is the volume of the Eulerian cell intersected. The governing parameters are chosen as $D_{\text{eq}} = 1$, $\Gamma = 2$, $\mathbf{g} = 2$, and $\nu$ is varied over the simulations using the values $\nu = \frac{1}{10},\ \nu = \frac{1}{50}$, or $\nu = \frac{1}{100}$.

We integrate equations (\ref{eq:ch3_NS}) relying direct numerical simulation with our in-house open-sourced advanced finite difference code
\href{http://www.afid.eu}{(``AFiD'')} which is extensively described and validated \citep{AFiD,VamRod}. The spatial derivatives are approximated by conservative,
second-order accurate finite-differences discretised on a staggered mesh which is uniform and homogeneous in all directions.
A combination of Crank--Nicolson and low-storage third-order Runge--Kutta schemes is used to integrate, respectively, the viscous terms implicitly and all other terms explicitly in time.
Finally, pressure and momentum are strongly coupled through a fractional-step method as described by \cite{MOHANRAI199115,VERZICCO1996402}.

In total, five numerical simulations of falling oloids were performed: one for $\nu = \frac{1}{100}$, and one for $\nu = \frac{1}{50}$, starting from the reference orientation shown in figure \ref{fig:Oloid_photo_model}c). The remaining three test cases were performed at $\nu = \frac{1}{10}$, with varying initial orientation. For all simulations, the oloid starts with zero linear and angular velocity, with the fluid at rest. The domain in which the oloid was simulated is a cube of $10 \ D_\text{eq} \times 10 \ D_\text{eq} \times 10$ $D_\text{eq}$, with periodic boundary conditions in all directions.

\section{Results}
\label{sec:Results}

\subsection{Settling velocity and different falling modes}
\begin{figure}
    \centering
    \includegraphics[width=\linewidth]{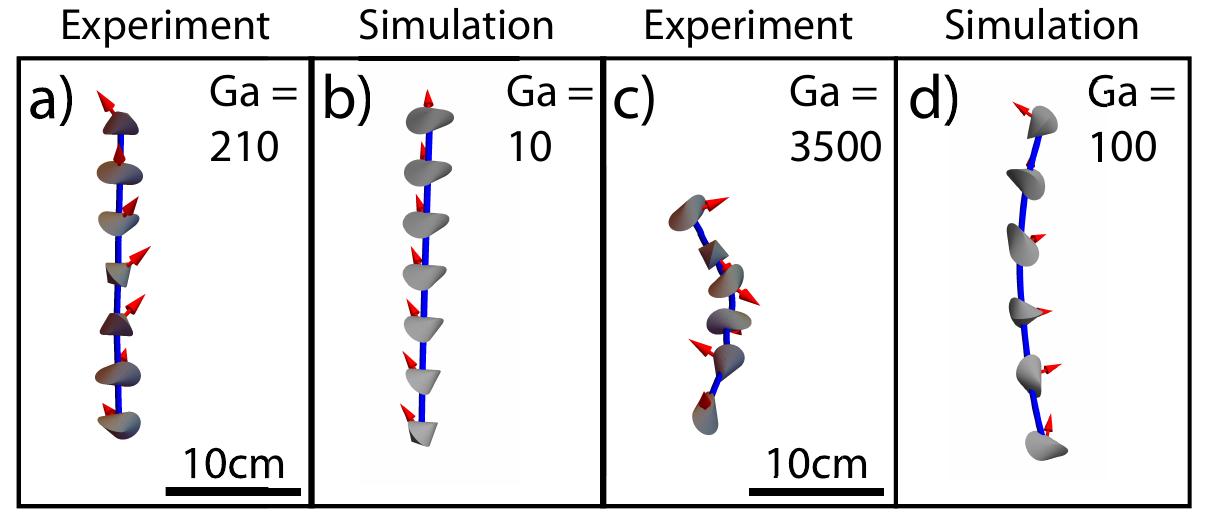}
    \caption{Snapshots of the oloid orientation for the two observed settling modes. The red arrow shows a pointing vector of the oloid, the blue line shows the path of the centre of mass. Snapshots of an experimentally measured oloid displaying a stable settling mode (a) for $\text{Ga} = 210$, with snapshots 0.6 s apart. Snapshots of a simulated oloid showing a stable settling mode (b) for $\text{Ga} = 10$. Snapshots of an experimentally observed tumbling oloid (c) for $\text{Ga} = 3500$, with snapshots 0.2 s apart. Snapshots of a simulated tumbling oloid (d) for $\text{Ga} = 100$.}
    \label{fig:oloid_settling_snapshots}
\end{figure}

The first observation from experiments and simulations is that the oloids fall in two distinct falling modes, depending on the Galileo number. At lower Galileo number ($\text{Ga} \leq 210$ for experiments, $\text{Ga} \leq 10$) the oloid settles in a stable manner, falling while preserving its orientation over time. For higher Galileo numbers ($\text{Ga} \geq 440$ in experiments, $\text{Ga} \geq 50$), the oloid tumbles as it settles: both falling modes were observed experimentally and numerically. Snapshots of a settling oloid in both regimes are shown in figure \ref{fig:oloid_settling_snapshots}. Figure \ref{fig:oloid_settling_snapshots}a) shows snapshots of an experimentally tracked oloid in the stable settling mode, where snapshots are shown for every 0.8 s. Figure \ref{fig:oloid_settling_snapshots}b) shows snapshots of the simulated oloid in the stable settling regime. Figure \ref{fig:oloid_settling_snapshots}c) shows snapshots of a tumbling oloid measured experimentally, with snapshots spaced by 0.2 s. Finally, figure \ref{fig:oloid_settling_snapshots}d) shows snapshots of the numerical simulation of a tumbling oloid. Animations of the simulations and videos of the experiments in both regimes can be found in the supplementary material.
The observed snapshots in both regimes are comparable between experiments and simulations, where in the stable regime the oloid rotates around the vertical axis. Comparing the experiments and simulations, we see a difference in the ratio of linear (downward) velocity to angular velocity, which we attribute to a difference in Galileo number. The oloid snapshots in the tumbling regime show a more chaotic motion. Due to the clear differences between the stable settling and the tumbling oloid, the subsequent results are categorised into these two regimes.

\begin{figure}
    \centering
    \includegraphics[width=0.8\linewidth]{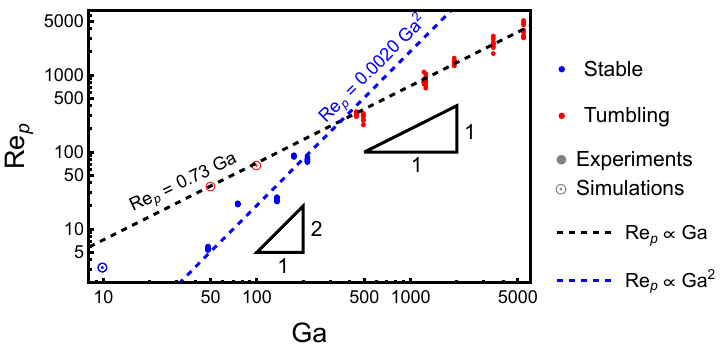}
    \caption{Particle Reynolds number as a function of the Galileo number. The blue points denote a stable settling oloid, whereas the red points are tumbling oloids. The black, dashed line shows $\text{Re}_p \propto \text{Ga}$ expected for turbulent drag. The blue, dashed line shows $\text{Re}_p \propto \text{Ga}^2$, which is the expected scaling for Stokes drag.}
    \label{fig:Re_Ga_plot}
\end{figure}

First, the average settling velocity for all oloids is determined per tracked oloid, and used to calculate the particle Reynolds number $\text{Re}_p = \frac{D_{\text{eq}} v_z}{\nu}$, with $v_z$ the average vertical velocity. Figure \ref{fig:Re_Ga_plot} shows the particle Reynolds number as a function of the Galileo number for the experimentally tracked oloids.
The blue and red points denote a stable falling oloid and a tumbling oloid, respectively. The black, dashed line indicates $\text{Re}_p \propto \text{Ga}$, where the proportionality factor was fitted to the data. The open circles with a dot represent the results from the simulations. We observe that the particle Reynolds number and Galileo number agree well for the tumbling oloids, and follow the linear scaling very well. Therefore, the estimated settling velocity $v_s = \sqrt{g D_{\text{eq}} \left(\frac{\rho_p}{\rho_f} - 1 \right)}$ as used in the Galileo number corresponds well to the measured settling velocity. Here the estimate for the velocity is derived from a force balance between gravity, buoyancy and drag, where the turbulent drag scales as $F_D \propto v^2$. The discrepancies between results from experiments and simulations are attributed to the difference in density ratio $\Gamma$. For the stable settling, the measured velocity is lower than the estimated settling velocity. We suspect that viscous effects dominate the dynamics in the stable settling regime, increasing the drag, and reducing the velocity. To verify this, we compute another estimated velocity, based on the viscous Stokes drag, which scales linearly with the particle velocity. In this regime where the Stokes drag applies, the particle Reynolds number then scales as $\text{Re}_p \propto \text{Ga}^2$, as shown in figure \ref{fig:Re_Ga_plot}. The blue, dashed line is fitted to the stable settling data with a quadratic scaling. Again, this scaling agrees well with the measurements of the stable settling oloids, indicating that the stable oloids are affected by Stokes drag. Therefore, in the Stokes regime, we find oloids settling in a stable manner, whereas in the tumbling regime, the oloids are beyond the Stokes regime and experience turbulent drag. In the tumbling regime, we again attribute the differences between experiments and simulations to the difference in density ratio.

\begin{figure}
    \centering
    \includegraphics[width=\linewidth]{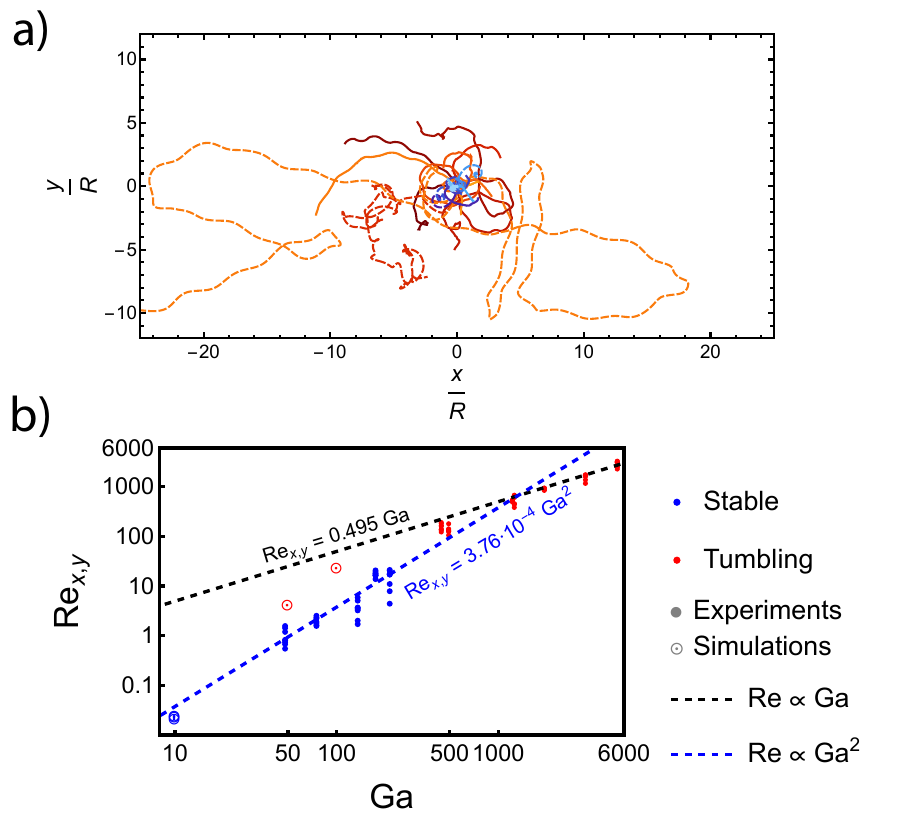}
    \caption{a) Some of the projected oloid trajectories, with blue and red indicating the stable settling and tumbling regime respectively. The dashed lines show trajectories from simulations over longer time scales compared to experiments. The trajectories are compensated to start at the origin. b) The Reynolds number based on the root mean square velocities in the $x,y$-plane plotted as a function of the Galileo number. Blue and red colours mark the stable settling and tumbling regime, respectively.}
    \label{fig:Trajectories}
\end{figure}

However, the trajectories of the oloids differ between the observed regimes, with the stable settling mode generally having trajectories with small displacements in the horizontal plane ($x$,$y$-plane), whereas the tumbling oloids show larger movements in those directions. This is already visible in figure \ref{fig:oloid_settling_snapshots}, but becomes more clearly visible in figure \ref{fig:Trajectories}a). 
This figure shows $20\%$ of the projected oloid trajectories from experiments, and all projected trajectories from simulations. The colour denotes the settling regime (blue shades for stable settling, red shades for tumbling). The solid lines show experimentally measured trajectories, whereas the dashed lines indicate trajectories from simulations. Here, we note that the simulations were performed over larger dimensionless timescales compared to experiments. The coordinate system has been chosen such that the trajectories start at the origin. Figure \ref{fig:Trajectories}a) clearly illustrates that oloids in the stable regime have smaller planar movements compared to the tumbling oloids. The projected trajectories mostly show curved paths, which occur both in the experimental cases and in the simulations. The oloids exhibit helical trajectories of both handednesses, mainly visible in the stable settling regime. The oloid's planar movement can be more quantitatively studied by computing the Reynolds number based on the planar velocity. Using the root mean square of the oloid velocities in the $x$,$y$-plane to compute the Reynolds number $\text{Re}_{x,y}$ gives the plot shown in figure \ref{fig:Trajectories}b). The dashed lines in this plot are fitted to the experimental data, with fixed exponents. Similar to the particle Reynolds number in figure \ref{fig:Re_Ga_plot}, the planar Reynolds number increases with Galileo number, and the results differ between the two regimes. The planar Reynolds numbers for the stable settling oloids are lower than for the tumbling oloids. This indicates slower planar movements for the stable settling regime, as expected from the shown projected trajectories. Again, differences between the results in experiments and simulations are attributed to the large difference in density ratio $\Gamma$.

So far, the results illustrate two different regimes, where the stable regime shows lower velocities in the vertical and horizontal directions compared to the tumbling regime. Additionally, the snapshots of the settling oloids in figure \ref{fig:oloid_settling_snapshots} seem to show different orientation dynamics between the two regimes. What remains is to describe these different falling modes in terms of the orientation and rotation of the settling particles, and compare the numerical results to the experimental findings more explicitly.

\subsection{Orientation data}
\begin{figure}
    \centering
    \includegraphics[width=\linewidth]{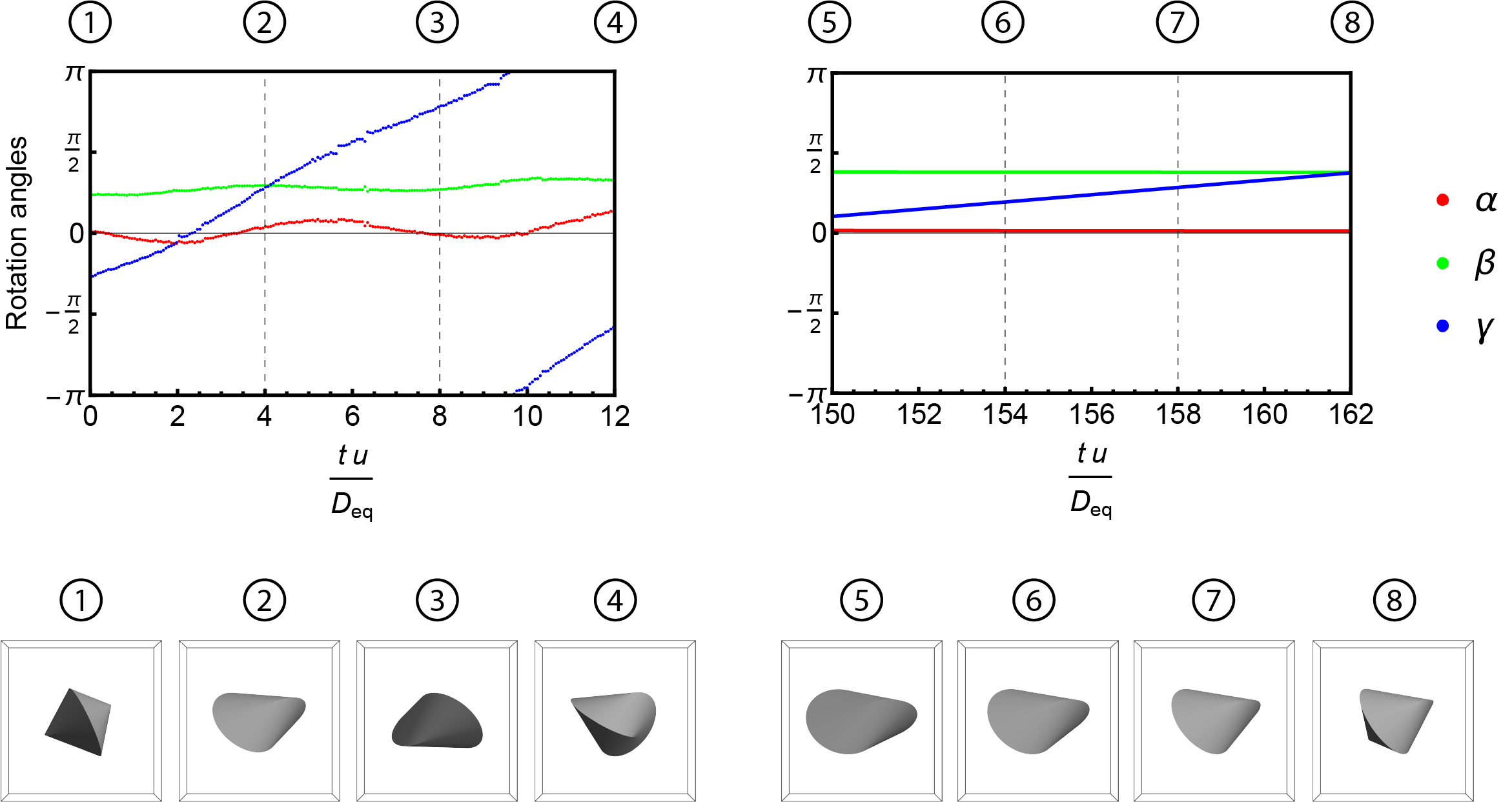}
    \caption{Rotation angles around the $x-y-z$ axes given by $\alpha$, $\beta$, and $\gamma$, respectively for a stable settling oloid. Experimentally tracked orientation angles (left panel), for the oloid shown in figure \ref{fig:oloid_settling_snapshots}a), where $\text{Ga} = 210$. Numerical orientation angles (right panel) for the oloid shown in figure \ref{fig:oloid_settling_snapshots}b), with $\text{Ga} = 10$. Snapshots of the oloid are shown at multiple nondimensional times to illustrate the particle dynamics.}
    \label{fig:Orientation angles stable}
\end{figure}

The tracked orientation of the oloid over time allows for a close study of its falling dynamics. First, the orientation of a stable settling oloid is compared between experiments and simulations, for which the settling results as in figure \ref{fig:oloid_settling_snapshots}a) and \ref{fig:oloid_settling_snapshots}b) are further investigated. Figure \ref{fig:Oloid_photo_model}c) shows the reference orientation for the oloid, and the pointing vector is represented by the red arrow. To find the orientation over time, we determine how the oloid is rotated with respect to the shown reference orientation.
The pitch, yaw, and roll angles (rotations with respect to the lab frame) over time are shown in figure \ref{fig:Orientation angles stable} along with orientation snapshots for both the experimental and numerical case, displayed in the left and right panel, respectively. The Galileo numbers in the experimental and numerical case are $\text{Ga} = 210$, and $\text{Ga} = 10$, respectively. The order of rotations is $x$-$y$-$z$, where $\alpha$ describes the rotation around the $x$-axis, $\beta$ around the $y$-axis, and $\gamma$ around the $z$-axis. For this figure, the experimental and numerical data are shown for an equal range of dimensionless time, to ensure a fair comparison. Here a later part of the simulation is shown, to ensure that the particle is at terminal velocity (like in the experimental results).
The rotation dynamics match well between experiments and simulations, and clearly indicate a rotation around the vertical axis, shown by the monotonously increasing angle $\gamma$, whereas the angles with respect to the $x$-axis and $y$-axis remain (approximately) the same. This confirms that the particles rotate around the vertical axis, while staying in an otherwise unchanged orientation. The rotation angle around the $x$-axis oscillates around $0$ in the experimental case, and stays at $0$ for the numerical case, indicating that the oloid's orientation remains unchanged with respect to the reference orientation around the $x$-axis.
The rotation around the $y$-axis, shown by $\beta$, is very similar in value in both the experimental and numerical case. This rotation angle now indicates the angle between the vertical axis and the pointing vector, visualised by the red vector in figure \ref{fig:Oloid_photo_model}c). Therefore, the `tilt angle' of the oloid with respect to the vertical axis remains constant during settling, while the particle rotates around the vertical axis. The rotation rate differs between the experiment and simulation, which we attribute to the difference in Galileo number.

\begin{figure}
    \centering
    \includegraphics[width=\linewidth]{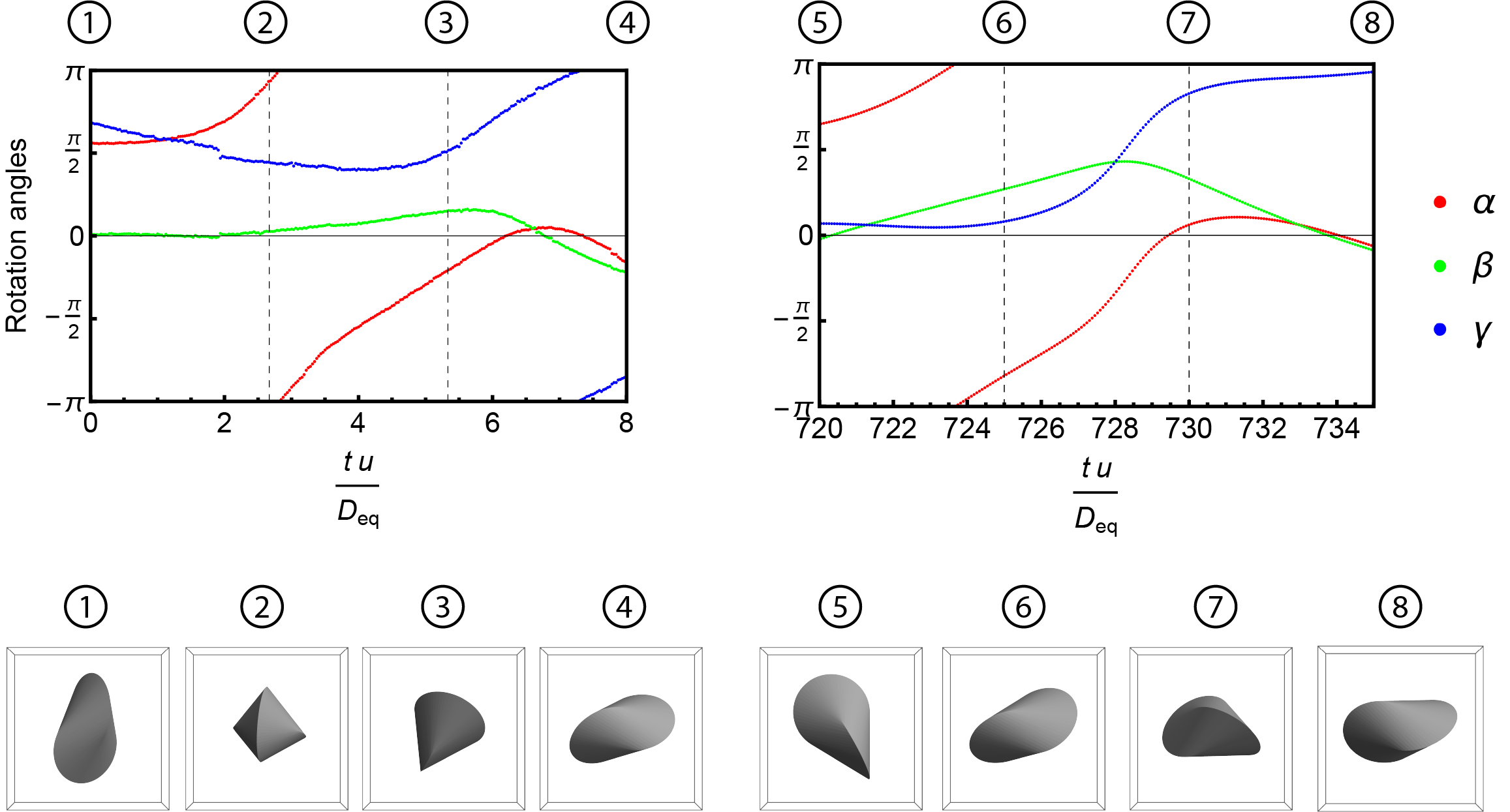}
    \caption{Rotation angles around the $x-y-z$ axes given by $\alpha$, $\beta$, and $\gamma$, respectively, for a tumbling oloid. Experimentally tracked orientation angles (left panel), for the oloid shown in figure \ref{fig:oloid_settling_snapshots}b), with $\text{Ga} = 1300$. Numerical orientation angles (right panel) for the oloid shown in figure \ref{fig:oloid_settling_snapshots}d), for $\text{Ga} = 100$. Snapshots of the oloid are shown at multiple nondimensional times to illustrate the particle dynamics.}
    \label{fig:Orientation angles tumbling}
\end{figure}

\begin{figure}
  \centering
  \includegraphics[width=\textwidth]{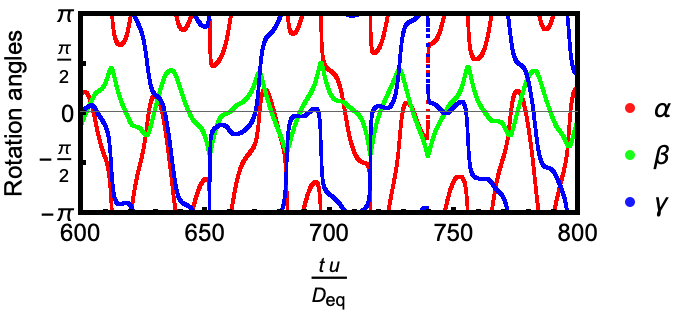}
  \caption{Rotation angles of the tumbling oloid over a longer observation time than figure \ref{fig:Orientation angles tumbling} (right panel), for $\text{Ga} = 100$.}
  \label{fig:Orientation angles long}
\end{figure}

The orientation over time for the tumbling oloid shown in figure \ref{fig:Orientation angles tumbling} shows very different dynamics. Again, the left panel shows the experimentally tracked rotation angles $\alpha$, $\beta$, $\gamma$. The right panel shows the same rotation angles for the numerical simulation of a tumbling oloid over a longer nondimensional time range. Both panels show snapshots of the oloid orientation at various dimensionless times to illustrate the particle dynamics.
Rather than a rotation around a single axis, all three rotation angles now change continuously, and the experimental and numerical results again match quite well qualitatively. Again, the time scales seem different between the experimental and numerical results: the dynamics in the experiments are similar, but happen over a shorter dimensionless time compared to the simulations, which we expect to be caused by a difference in Galileo number. The simulations allow us to study the rotation angles over a longer time period, as shown in figure \ref{fig:Orientation angles long}, showing the rotation angles as displayed in figure \ref{fig:Orientation angles tumbling} over a longer dimensionless timescale.
This figure indicates that there are spans of quasi-periodic motion, though the periodicity is broken over longer timespans. Closely inspecting this figure also shows that the orientation dynamics in figure \ref{fig:Orientation angles tumbling} closely resemble those in the simulation around $720 \leq \frac{t u}{D_{\text{eq}}} \leq 735$.

Overall, the results from the experiments and simulations qualitatively agree, and show that the numerical model is capable of capturing the general trends of the oloid settling dynamics. We now move away from analysing single oloid measurements, and analyse the data gathered from all settling oloids as a whole, in order to gain a better understanding of the full settling process.

\begin{figure}
    \centering
    \includegraphics[width=\textwidth]{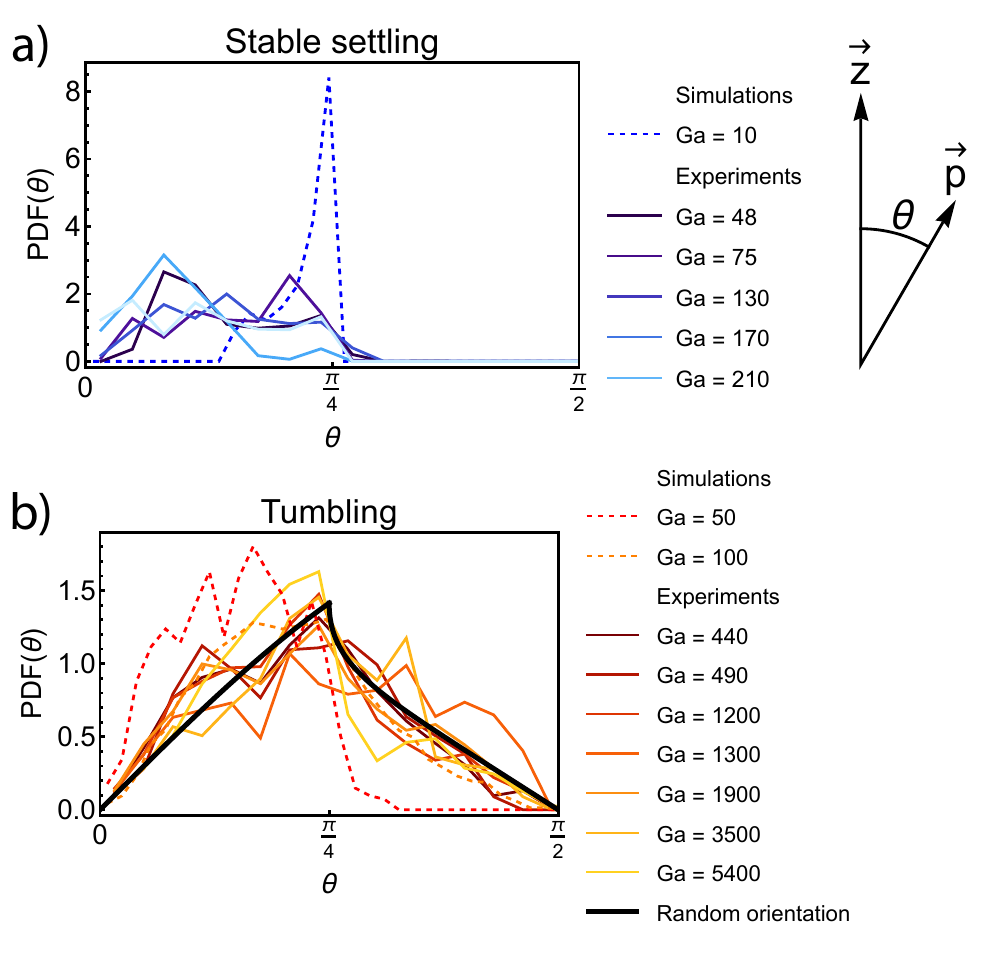}
    \caption{The PDFs of the angle between the oloid's pointing vector and the vertical for the stable settling oloids a), and the tumbling regime b). The fourfold symmetry of the oloid is taken into account, by taking the minimum of the angles of equivalent orientations. The definition of the angle between the vertical and the pointing vector is shown in the top-right graphic. The PDF of the angle between the vertical and the pointing vector of a randomly-oriented oloid is analytically found in Appendix \ref{app:Oloid distribution}.}
    \label{fig:upvecpdf}
\end{figure}

For this purpose, we compute the angle of the oloid pointing vector with respect to the vertical axis for each time step (both experimentally and numerically), for all tracked oloids. Here the fourfold symmetry of the oloid is taken into account to compensate for any equivalent orientations, taking the minimum angle of the equivalent orientations. To ensure a fair comparison, only the numerical data is used after the oloid has achieved a vertical velocity of $99\%$ of the terminal velocity. These results are displayed in figure \ref{fig:upvecpdf}, where the graphic in the top-right defines the angle between the pointing vector and the vertical axis. For each Galileo number, all tracked oloids are grouped together for the PDFs.
Figure \ref{fig:upvecpdf}a) shows the PDFs of the angle between the oloid pointing vector and the vertical for the stable settling regime. The most striking feature of this plot is that the PDFs in the stable settling regime are almost completely situated in the range $0 \leq \theta \leq \frac{\pi}{4}$, whereas the other half of the (possible) range is empty. This indicates that the oloid never tips end-over-end, and always remains in an orientation similar to the ones seen in figures \ref{fig:oloid_settling_snapshots}a) and \ref{fig:oloid_settling_snapshots}b), where its pointing vector has a tilting angle with respect to the vertical. Therefore, despite the random initial orientations of the oloid, it has a clear preferential orientation range. This is emphasised by the supplementary material: one of the videos shows an oloid achieving its preferential orientation quickly after release. The exact angle between the oloid pointing vector and the vertical differs between experiments, as shown by the wide PDFs found from the experimental data.

The numerical results, shown by the dashed line (combining the data for all three simulations at $Ga = 10$), do show a clear single peak, indicating a strong preferential orientation. In the numerical simulations, $\theta$ attains the same value ($\theta = \pi/4$) for all three simulations at $Ga = 10$, regardless of the oloid initial orientation. This differs from the experimental findings, which show a wide range of possible $\theta$ values. A possible explanation for this discrepancy is the initial (angular) velocity of the oloid in experiments, which was difficult to control due to the way the oloids were released in the fluid. Additionally, small errors in the 3D-printed oloids result in not perfectly symmetric particles due to small defects. Besides this, the fluid in the experimental tank will always be slightly in motion due to the release platform displacing fluid when the oloid is released. This small flow may also affect the orientation dynamics at these low Galileo numbers. These explanations remain speculative however, since these initial conditions were not varied in the simulations, and remain outside the scope of this study.

Overall, the experiments and numerical simulations do show a clear trend in the stable oloid settling, where the oloids orient themselves with a tilt with respect to the vertical. The magnitude of this tilt angle may depend on initial conditions, which was not controlled for in experiments.

Similar statistics can be obtained for the tumbling regime, for which the PDFs of the angle between the oloid's pointing vector and the vertical are displayed in figure \ref{fig:upvecpdf}b). Comparing this to the stable settling regime, it is immediately clear that the tumbling oloids do not display a clear preferential orientation, which was the case for the stable settling oloids. For the tumbling oloids, the orientation seems randomly distributed, which can be verified by plotting the PDF of the angle $\theta$ for random orientations, indicated by the thick black line. The expression for this distribution can be found analytically, which is shown in Appendix \ref{app:Oloid distribution}. The shape of the random distribution has an asymmetric shape due to the oloid's shape: the oloid has a twofold symmetry when rotating around the $y$-axis, but has no such symmetry around its other axes. Additionally, the oloid has reflective symmetry in the $x$-$y$ and $y$-$z$ planes. This results in a difference in the maximum angle between the vertical and the pointing vector, when comparing a rotation around the $x$-axis and $y$-axis. As a result, the PDF for the random orientations is asymmetrical in $\theta$. Overall, the experimental results show a clear transition towards a random distribution as the Galileo number increases. For the lowest Galileo numbers, the PDF differs significantly from the random distributions, indicating the existence of a transition between the stable and tumbling regime. This is further supported by the numerical results, with the higher Galileo number agreeing very well to the experimental findings in the transitional regime. The simulation at $\text{Ga} = 50$ shows completely different results compared to the other PDFs in the figure. We attribute this to the much lower Galileo number, which is close to the stable regime. Therefore, the PDF of the simulation of lower $\text{Ga}$ is similar to the stable settling PDFs, indicating this simulation is in the transitional regime, close to the stable regime. These curves together indicate that, as the Galileo number increases, an oloid transitions from a preferential orientation towards a random orientation. In between the stable and the tumbling regime, a transitional regime may be identified.

\subsection{Rotation data}
\begin{figure}
    \centering
    \includegraphics[width=\textwidth]{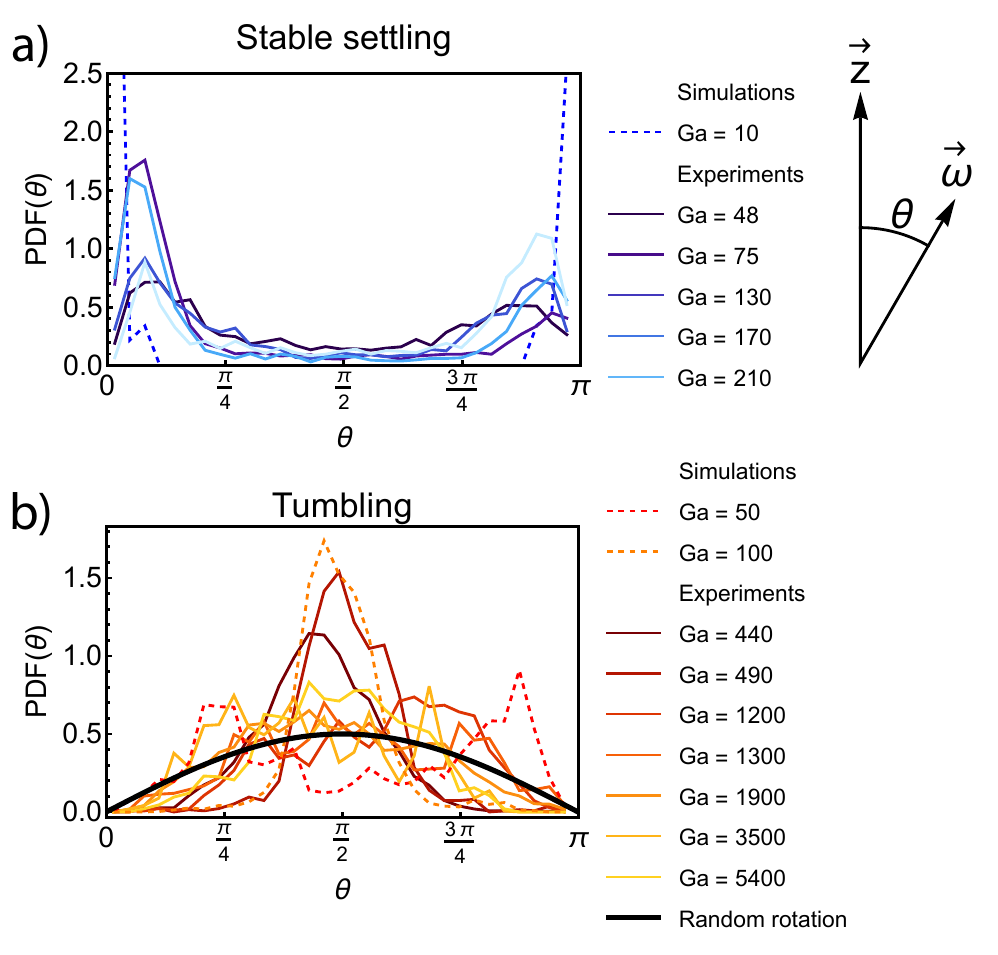}
    \caption{The PDFs of the angle between the oloid's rotation vector and the vertical, for the stable settling oloids a) and the tumbling oloids b). The rightmost graphic shows the definition of the angle between the rotation vector and the vertical.}
    \label{fig:rotvecpdf}
\end{figure}

In addition to the orientation statistics of the oloid, the rotation statistics are investigated and compared between the two settling regimes. Similar to the orientation data, the rotation of the oloid is studied by computing the rotation vector and finding its angle with respect to the vertical. Note that the rotation vector is unaffected by the oloid's symmetry, and by the choice of pointing vector. Figure \ref{fig:rotvecpdf}a) shows the PDFs of the angle between the oloid's rotation vector and the vertical for the stable settling regime.
This figure shows a clear bimodal distribution of the rotation vector around $\theta = 0$ and $\theta = \pi$ for all Galileo numbers in the stable settling regime, both for the experimental and numerical results. This corresponds to a rotation about the vertical axis, where both rotation directions occur. This confirms the findings in the previous section, where the stable settling oloids were shown to rotate around the vertical axis when settling. This figure additionally shows that the oloid can rotate in either direction around the vertical axis, which occurs both in experiments and simulations for all tested Galileo numbers in the stable settling regime.

For the tumbling regime, figure \ref{fig:rotvecpdf}b) shows the PDFs of the angle between the oloid's rotation vector and the vertical. Similar to the orientation data in this regime, the rotation data for tumbling strongly differs from the stable regime. The rotation vector seems randomly distributed, which is evidenced by the thick black line, showing the PDF for a random vector distribution. This random distribution corresponds to a $\sin({\theta})/2$ curve, as can be found theoretically. The lowest experimental Galileo numbers in the tumbling regime show PDFs which deviate from the random distribution, and show a peak near $\theta = \pi/2$. This again indicates a transition between the stable and tumbling regimes, which is supported by the simulation results. The simulation at $\text{Ga} = 50$ shows two peaks on either side of the centre, similar (though less pronounced) to what is observed for the stable settling regimes. The simulation at $\text{Ga} = 100$ again agrees very well with the experimental findings in the transitional regime. Here the Galileo numbers where the transition between stable and tumbling occurs differ between experiments and simulations, which we attribute to the large difference in $\Gamma = \frac{\rho_p}{\rho_f}$ between experiments and simulations. Overall, we find a transition from a rotation around the vertical axis to a random rotation as the Galileo number increases, similar to the observations made for the pointing vector alignment.\\

\begin{figure}[ht!]
  \centering
  \includegraphics[width=\textwidth]{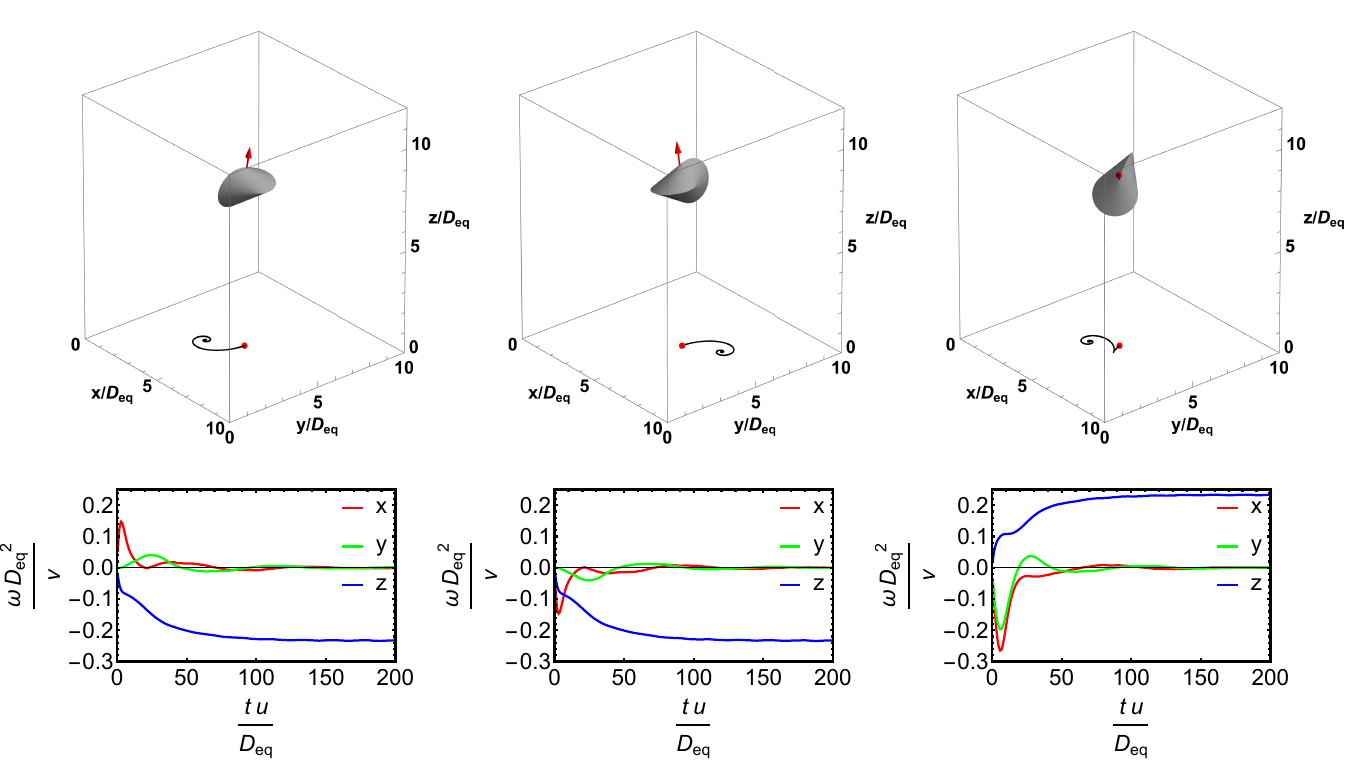}
  \caption{The first row depicts the initial orientation of the oloid for three different cases at a low Galileo number ($\text{Ga} = 10$), where the centre of mass trajectory is projected on the bottom plane. The red vector shows the pointing vector, and the red dot shows the initial projected centre of mass position. The second row displays the time evolution of the angular velocity components with respect to the lab frame.}
  \label{fig:Initial orientation effects}
\end{figure}

The occurrence of a rotation around the vertical axis in both directions in the stable regime is striking, since the particle itself is non-chiral. Here a parallel can be drawn to the oblique settling mode found in spheres, reported by \cite{Uhlmann2014}. This oblique regime also breaks symmetry, without the particle breaking symmetry. In the current case of the oloid, the particle shape does not break mirror symmetry, hence the cause for the observed translation-rotation coupling must be found elsewhere. To this extent, the three numerical simulations in the stable regime are compared to each other, where the starting orientation is varied.
The effects of the initial orientation are shown in figure \ref{fig:Initial orientation effects}. The top row shows the initial orientation of the oloid and illustrates the trajectory of the centre of mass projected on the bottom plane. Here the red arrow shows the oloid's pointing vector, whereas the red dot on the bottom plane indicates the initial position of the centre of mass. The second row shows the angular velocity components over time in the lab coordinate frame.
This figure shows that the angular velocity components vary significantly with the initial orientation.
Focusing on the first two columns, where the second column shows an initial orientation which is simply a 180° rotation about the $x$-axis (or a 180° rotation about the $z$-axis) with respect to the initial orientation in the first column, we observe no difference, as expected by symmetry. In contrast, when the oloid starts with a tilted orientation (third column), the $z$-component of the angular velocity has now flipped sign when looking at the later timescales, indicating an opposing rotation direction compared to the first two columns. The animation of the simulations at low Galileo number in the supplemental material clearly shows this result. Therefore, the initial orientation affects the rotation direction of the settling oloid, similar to results by \cite{Miara2024}, who show that U-shaped disks can settle in helical trajectories, where the chirality of the trajectory is determined by the disk's initial orientation.

These findings highlight that the oloid dynamics are highly sensitive to its initial orientation. This sensitivity has important implications, particularly in experimental setups where it is difficult to precisely control the initial orientation. Further initial conditions or slight geometric asymmetries may therefore explain the difference between the numerical and experimental results shown in figure \ref{fig:upvecpdf}, and be the cause for a wide range of observed tilt angles observed in the stable settling regime. Hence, extra numerical simulations could be performed to explore additional initial conditions, which may affect the oloid's settling dynamics in ways that cannot be easily predicted. 
Among others, a parallel can be drawn with Rayleigh--Bénard convection \citep{rayleigh-benard}, where the system achieves a steady state, but the specifics of whether the bulk flow (``the wind") rotates clockwise or counter-clockwise depend on subtle variations in the initial conditions. While the macroscopic behaviour is consistent across experiments, the microscopic details of the flow are sensitive to initial perturbations.

\begin{figure}
    \centering
    \includegraphics[width=\linewidth]{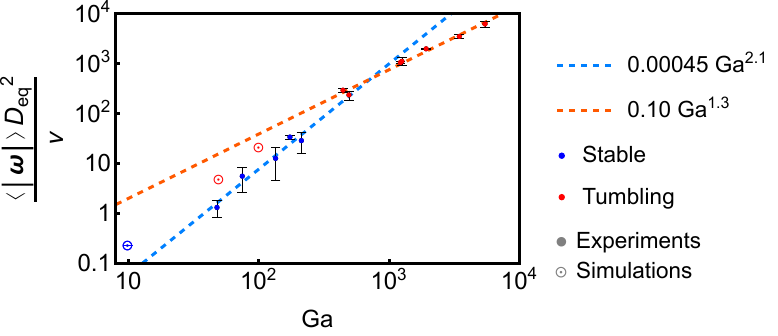}
    \caption{Average magnitude of the angular velocity as function of the Galileo number. The angular velocity is nondimensionalised by $D_{\text{eq}}$ and $\nu$. The error bars show a standard deviation between tracks of the same Galileo number. The blue and red dashed lines represent best fits to the experimental stable data and experimental tumbling data, respectively.}
    \label{fig:Omega_Ga_plot}
\end{figure}

Finally, we investigate how the angular velocity scales with the Galileo number, which is shown in figure \ref{fig:Omega_Ga_plot}.
In this figure, $\langle |\boldsymbol{\omega}| \rangle$ is the mean absolute value of the angular velocity, where the mean absolute angular velocity is computed per particle track, and then averaged over all tracks using a weighing by track length. The mean absolute angular velocity is made dimensionless using $D_{\text{eq}}$ and $\nu$. The error bars in the figure indicate the standard deviation of mean angular velocity between tracks of the same Galileo number.

The graph in figure \ref{fig:Omega_Ga_plot} shows a clear increase of angular velocity with the Galileo number, where the blue and red dashed lines show fits to the experimental data in the stable and tumbling regimes, respectively. The best fits are close to the expected Reynolds number scaling as seen in \ref{fig:Re_Ga_plot}, which may be expected, since the dimensionless group $\frac{\langle |\boldsymbol{\omega}| \rangle D_{eq}^2}{\nu}$ is a de facto Reynolds number. The figure highlights that next to the linear velocity, the angular velocity scaling in the stable regime is also dictated by the Stokes drag. This emphasises once again that the stable and tumbling regime are separated by whether or not the settling oloid is in the Stokes regime.

\section{Conclusions and outlook}
\label{sec:Outlook and conclusions}
In this paper, we have conducted a combined experimental and numerical study of the settling dynamics of an oloid-shaped particle in a quiescent fluid. Our results identify two clear settling regimes: a stable settling regime at low Galileo numbers ($\text{Ga}\leq 210$), and a tumbling regime at higher Galileo numbers ($\text{Ga} \geq 440$). The stable regime is characterised by a translation rotation-coupling, where the oloid rotates around the vertical axis as it descends, while staying in a tilted orientation with respect to the vertical. In this regime, the drag is given by the Stokes drag, and determines the settling velocity. The tumbling regime, on the other hand, shows no preferential orientation or rotation. Our results show evidence of a gradual transition between these regimes, observed in the experiments and simulations. The experiments and simulations align well qualitatively, though a larger range of tilt angles in the stable regime is found experimentally compared to numerical results. This may be due to the difference in Galileo number, or due to any slight asymmetries in the particle geometry.

Numerical simulations reveal that the oloid angular velocity with respect to the vertical axis changes sign depending on the starting orientation. This shows that symmetry-breaking motion can be achieved without the particle geometry breaking mirror symmetry, as was shown in works by others \citep{Miara2024}.
Other variations in initial conditions can be investigated numerically, which may resolve any discrepancies between experiments and simulations.

The large discrepancy between Galileo numbers in the experiments and simulations calls for extra investigation, to unify the findings between experiments and simulations. The density ratio $\Gamma$ currently differs significantly between the simulations and experiments, hence unifying this parameter may resolve some of the mentioned discrepancies. Additionally, simulations of an oloid in the tumbling regime would be valuable for unifying the experimental results with simulations in this regime. Longer experimental particle tracks would shed light on whether any periodic dynamics occur for tumbling oloids, as suggested by the simulation. Additional experiments could also be performed to prove the existence of a transitional regime between the stable and tumbling regimes.

Overall, this study enhances our understanding of the dynamics of anisotropic particles like the oloid, and sets the stage for future research. Extending the range of numerical simulations and incorporating a broader range of initial conditions will provide deeper insights into the interplay between shape, orientation, and fluid-structure interactions in low Reynolds number regimes.

%
 \bmhead[Acknowledgements]{The authors thank Gert-Wim Bruggert, Martin Bos, and Thomas Zijlstra for their technical support. We would like to thank Federico Toschi for insightful discussions.}
 \bmhead[Funding]{The authors acknowledge the access to several computational resources for this study: the European High Performance Computing Joint Undertaking for awarding us access to Discoverer under the project EHPC-REG-2022R03-208, and the national e-infrastructure of SURFsara, a subsidiary of SURF cooperation. This research has been funded by the Dutch Research Council (NWO) under grant OCENW.GROOT.2019.031.}
 \bmhead[Declaration of interests]{The authors report no conflict of interest.}
\bmhead[Data availability statement]{The raw data are available upon reasonable request.}
%
%

\appendix
\begin{appen}

\section{Random orientation distribution}\label{app:Oloid distribution}
Due to the symmetries of the oloid shape, the distribution of the pointing vector is nontrivial for a randomly oriented oloid. Figure \ref{fig:upvecpdf} shows the distribution of the angle between the pointing vector and the vertical for a randomly oriented oloid. The shape of this distribution is nontrivial, but can be analytically derived. We start with the oloid and its pointing vector in a random orientation: given the symmetries of the oloid, we can define 4 equivalent pointing vectors, as indicated by the red arrows in figure \ref{fig:oloidaxes}. For the angle between the pointing vector and the vertical, we take the smallest of the 4 angles between the pointing vectors (red arrows) and the vertical (black arrow).

\begin{figure}[htbp!]
    \centering
    \includegraphics[width=0.35\linewidth]{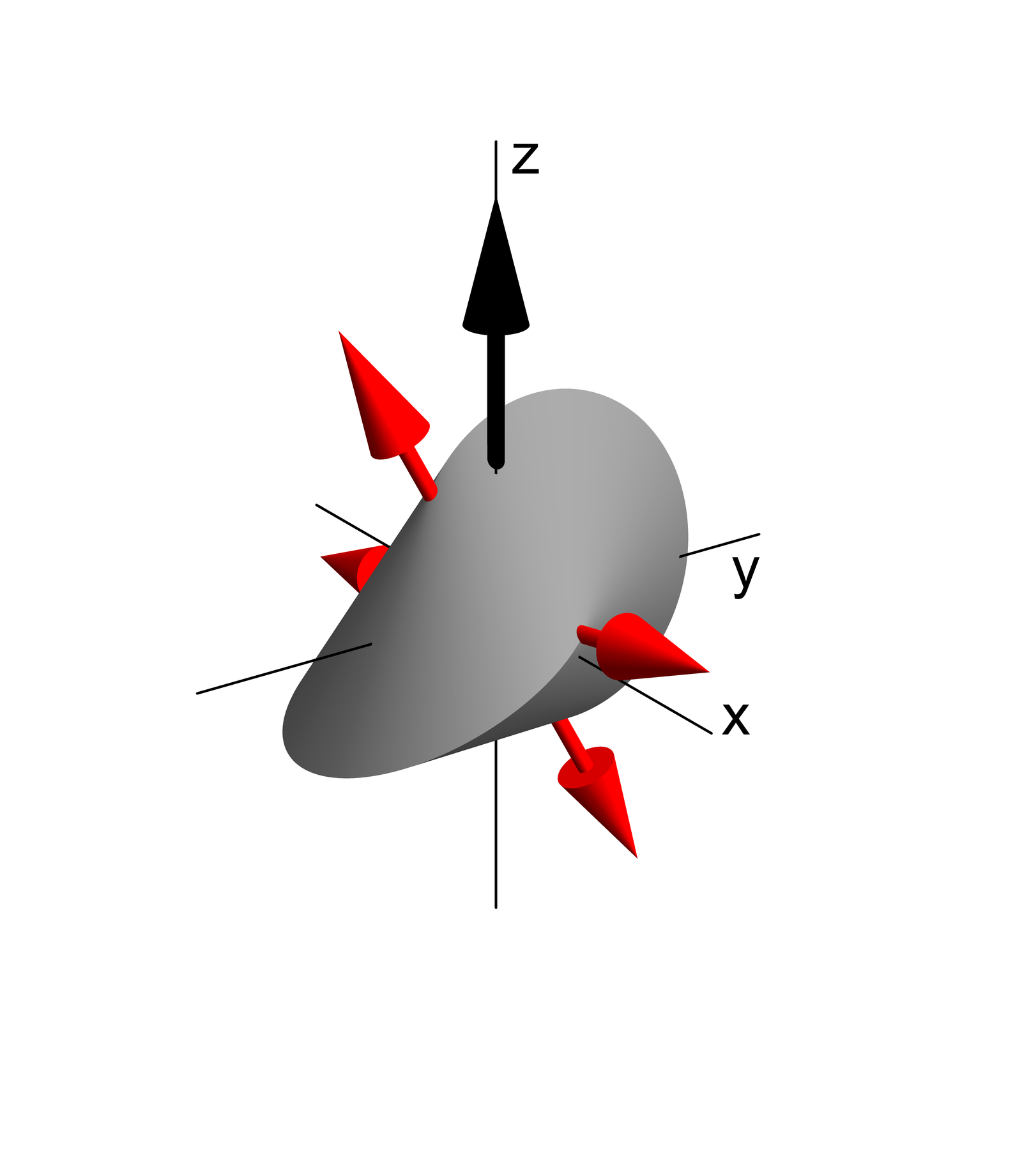}
    \caption{Pointing vectors of the oloid particles, taking into account the symmetries of the shape.}
    \label{fig:oloidaxes}
\end{figure}

To obtain the random orientations of the oloid, we use Shoemake's method \citep{Shoemake1992} to generate uniform random unit quaternions $\vec Q$, generating uniform random orientations: 
\begin{align}
    u & \dist \uniform{0}{1} \label{eq:algo1}\\
    \theta_1 & \dist \uniform{0}{2\pi} \\
    \theta_2 & \dist \uniform{0}{2\pi} \\
    r_1 &= \sqrt{1-u}\\
    r_2 &= \sqrt{u}\\
    \vec Q &= \begin{bmatrix}
\sin(\theta_1) r_1 \\
\cos(\theta_1) r_1 \\
\sin(\theta_2) r_2 \\
\cos(\theta_2) r_2
\end{bmatrix}\label{eq:algo6}
\end{align}
where $\dist$ means ``distributed as'' and $\uniform{A}{B}$ represents a uniform distribution between $A$ and $B$. This quaternion $\vec Q = [x,y,z,w]$ can be written as a rotation matrix $\vvec{R}$:
\begin{align}
\vvec{R} &= \begin{bmatrix}
 1-2 \left(y^2+z^2\right) & 2 (x y-w z) & 2 (w y+x z) \\
 2 (w z+x y) & 1-2 \left(x^2+z^2\right) & 2 (y z-w x) \\
 2 (x z-w y) & 2 (w x+y z) & 1-2 \left(x^2+y^2\right) \\
\end{bmatrix}
\end{align}
The 4 red arrows (pointing vectors) are given by:
\begin{align}
\vec{v_1} = \begin{bmatrix}
0 \\ 0 \\ 1
\end{bmatrix},
\vec{v_2} = \begin{bmatrix}
0 \\ 0 \\ -1
\end{bmatrix},
\vec{v_3} = \begin{bmatrix}
0 \\ 1 \\ 0
\end{bmatrix},
\vec{v_4} = \begin{bmatrix}
 0 \\ -1 \\ 0
\end{bmatrix}.
\end{align}
The angle $\theta$ between two unit vectors $\vec v$ and $\vec w$ is given by:
\begin{align}
    \theta &= \acos(\vec v \cdot \vec w)
\end{align}
The minimum angle between the rotated pointing vectors and the vertical ($\theta$) is then given by
\begin{align}
    \theta &= \min(\acos((\vvec{R} \cdot  \vec v_i)\cdot \vec Z))
\end{align}
where $\vec Z = [0,0,1]$ is the vertical, and $i$ goes from $1$ to $4$. Since $\vec Z$ is only non-zero in the last component, only the last row of $\vvec{R} \cdot  \vec v_i$ is important. Since $\vec v_i$ is always zero in the first component, this means only the last two columns of $\vvec{R}$ are needed. Combining this, only the following terms remain:
\begin{align}
    \theta &= \min(\acos(\begin{bmatrix} 2 (w x+y z) \\ 1-2 \left(x^2+y^2\right)\end{bmatrix} \cdot \vec{\tilde{v_i}}))
\end{align}
where $\vec{\tilde{v_i}}$ are the last two components of $\vec{v_i}$. Summarizing and redefining our symbols:
\begin{align}
    u & \dist \uniform{0}{1} \\
    \theta_1 & \dist \uniform{0}{2\pi} \\
    \theta_2 & \dist \uniform{0}{2\pi} \\
    r_1 &= \sqrt{1-u}\\
    r_2 &= \sqrt{u}\\
    \vec{v_1} &= \begin{bmatrix}
    0 \\ 1
    \end{bmatrix},
    \vec{v_2} = \begin{bmatrix}
    0 \\ -1
    \end{bmatrix},
    \vec{v_3} = \begin{bmatrix}
    1 \\ 0
    \end{bmatrix},
    \vec{v_4} = \begin{bmatrix}
    -1 \\ 0
    \end{bmatrix}\\
    \theta &= \min\left(\acos\left(\begin{bmatrix}
        -2 r_1 r_2 \sin(\theta_1-\theta_2) \\
        1- 2 r_1^2
    \end{bmatrix}\cdot v_i\right)\right)
\end{align}
Since $\acos$ is monotonically decreasing $\min \circ \acos$ = $\acos \circ \max$;
\begin{align}
    \theta &= \acos\left(\max\left(\begin{bmatrix}
        -2 r_1 r_2 \sin(\theta_1-\theta_2) \\
        1- 2 r_1^2
    \end{bmatrix}\cdot v_i\right)\right)
\end{align}
Taking the dot product with $v_i$ is observed to simply flip the sign of the remaining terms of the rotation matrix. Since we are looking for the largest number as argument for the $\acos$, we take the absolute value, as this gives the maximum of the multiplication options $v_1$ and $v_2$ or $v_3$ and $v_4$. Furthermore, we replace $r_1$ and $r_2$ by their definitions to obtain
\begin{align}
    u & \dist \uniform{0}{1} \\
    \Delta\theta & \dist \triangledist{-2\pi}{2\pi} \\
    \theta &= \acos\left(\max\left(
        2\sqrt{u(1-u)} \left|\sin(\Delta\theta)\right|,
        \left| 2 u -1 \right|
    \right)\right) \label{eq:depend2}
\end{align}
where $\triangledist{A}{B}$ is a triangular distribution from $A$ to $B$ (and having a mode (peak) at $(A+B)/2$ unless specified otherwise), and $\left| X \right|$ is the absolute value of $X$. We rewrite the $2\sqrt{u(1-u)}$ term in equation \eqref{eq:depend2} to simplify the problem further:
\begin{align}
    2\sqrt{u(1-u)} &= \sqrt{4u(1-u)} = \sqrt{4u-4u^2} \\
    &= \sqrt{1-(2u-1)^2}
\end{align}
In this form, we have $v=2u-1 \dist \uniform{-1}{1}$, since $ u \dist \uniform{0}{1}$. Applying the map $v \rightarrow \sqrt{1-v^2}$ depends only on $v^2$, thereby folding the domain $[-1,1]$ to $[0,1]$: therefore, the function $\sqrt{1-u^2}$ gives the same result. We have therefore performed the substitution $2u-1 \rightarrow u$, and we apply this substitution to both terms in equation \ref{eq:depend2} (since the two terms in the max are not independent), such that the function $g=\left| 2 u -1 \right|$ becomes $g= u $. The applied substitution then rewrites the problem statement to the form
\begin{align}
    u & \dist \uniform{0}{1} \\
    \Delta\theta & \dist \triangledist{-2\pi}{2\pi} \\
    \theta &= \acos\left(\max\left(
        \sqrt{1-u^2} \left|\sin(\Delta\theta)\right|,
        u
    \right)\right) \label{eq:theta_u}
\end{align}

To find the distribution of $\theta$, the distributions for each term in eq. \eqref{eq:theta_u} are found separately, and are later combined to give the distribution for $\theta$.

Considering that $u \in [0,1]$ , this implies $Y=\sqrt{1-u^2} \in [0,1]$, so only $0 \leq y \leq 1$ needs to be considered. This gives the CDF $F_Y(y)$ of the transformed function:
\begin{align}
 F_Y(y) &= P(\sqrt{1-u^2} \leq y) \\
 &= P(1-u^2 \leq y^2) \label{eq:ineq}
\end{align}
Solving this equation gives a single boundary point on the considered domain: 
\begin{equation}
    u_1 = \sqrt{1-y^2}
\end{equation}
Since $\sqrt{1-u^2}$ is monotonically decreasing between 0 and 1, inequality eq.\ \ref{eq:ineq} is true over the interval $[u_1,1]$. As $u$ is uniform and equal to 1, the probability is given by the length of the interval:
\begin{align}
    F_Y(y) &= P(u \in [u_1,1]) = 1-u_1\\
    &= 1 - \sqrt{1-y^2}
\end{align}
We differentiate the CDF to obtain the PDF ($f_Y(y)$):
\begin{align}
    f_Y(y) &= \frac{d}{dy}F_Y(y) = \frac{y}{\sqrt{1-y^2}}
\end{align}
on the domain $0\leq y \leq 1$ and $0$ otherwise.

Next, we find the distribution of the function $|\sin(\Delta \theta)|$: considering $|\sin(x)|$ is even, we can fold the triangular distribution and redefine our symbols accordingly:
\begin{align}
    \Delta\theta & \dist \triangledistmode{0}{2\pi}{0} \\
    f_{\Delta\theta}(t) &= \begin{cases}
 \frac{2 \pi -t}{2 \pi ^2} & 0\leq t\leq 2 \pi \\
 0 & \text{otherwise}
\end{cases}
\end{align}
We now evaluate the PDF of $\left|\sin(\Delta\theta)\right|$, in a similar manner as for the $\sqrt{1-u^2}$ function. Again, the function has a range $0 \leq y \leq 1$. The first solution (boundary point) is found as follows:
\begin{align}
    \tau^* = \asin(y) & & \text{for $0\leq\tau^*\leq\pi/2$}
\end{align}
\begin{figure}[htbp!]
    \centering
    \includegraphics[width=0.65\linewidth]{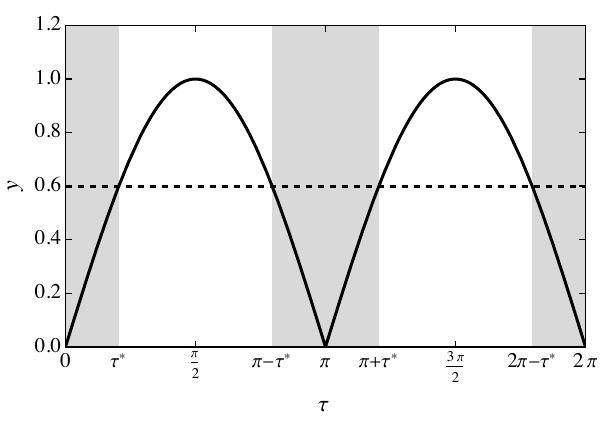}
    \caption{Domain interval of $\tau$ for $y<|\sin(\tau)|$.}
    \label{fig:abssindomain}
\end{figure}

All boundary points can be found, which read (see also figure \ref{fig:abssindomain}):
\begin{align}
    \tau &= \tau^*\\
    \tau &= \pi - \tau^*\\
    \tau &= \pi + \tau^*\\
    \tau &= 2\pi - \tau^*
\end{align}
These boundary points give the intervals
\begin{align}
    [0,\tau^*] \cup [\pi-\tau^*,\pi+\tau^*] \cup [2\pi-\tau^*,2\pi]
\end{align}
as sketched in figure \ref{fig:abssindomain}. Integrating the PDF over the intervals we obtain:
\begin{align}
F_U(y) &= \frac{1}{2\pi^2} \left(\int_{0}^{\tau^*} (2\pi - \tau)\,d\tau + \int_{\pi - \tau^*}^{\pi + \tau^*} (2\pi - \tau)\,d\tau +  \int_{2\pi - \tau^*}^{2\pi} (2\pi - \tau)\,d\tau \right) \\
 F_U(y) &= \frac{2\tau^*}{\pi} = \frac 2\pi \asin(y)
\end{align}
The PDF is then found by differentiating $F_U(u)$:
\begin{align}
    f_U(u) &= \frac d{du} F_U(u) = \frac{2}{\pi \sqrt{1-u^2}}.
\end{align}
The previously found distributions then rewrite the problem statement to
\begin{align}
    u & \dist \uniform{0}{1} \\
    v & \dist \frac{2}{\pi \sqrt{1-t^2}} \\
    \theta &= \acos\left(\max\left(
        \sqrt{1-u^2} v,
       u
    \right)\right) \label{eq:theta_uv}
\end{align}
From the terms inside the $\max$ we construct a joint PDF. Since these axes are not independent and $u \in [0,1]$ we find that this joint PDF fills only the first quadrant of the unit circle. The domain of the distribution of $v$ is compressed by $\sqrt{1-u^2}$ (and it's magnitude scaled by $1/\sqrt{1-u^2}$) for a value of $u$. The joint PDF is then given by:
\begin{align}
    f(u,t) &= \frac{2}{\pi \sqrt{1-\left(\frac{t}{\sqrt{1-u^2}}\right)^2}} \frac 1{ \sqrt{1-u^2}} \\
    f(u,t) &= \frac{2}{\pi \sqrt{1-t^2-u^2}}
\end{align}
This joint PDF is only positive in the first quadrant of the unit circle, hence we consider the output of the $\max$ function in this domain, as displayed in figure \ref{fig:maxvalue}.
\begin{figure}[htbp!]
    \centering
    \includegraphics[width=0.5\linewidth]{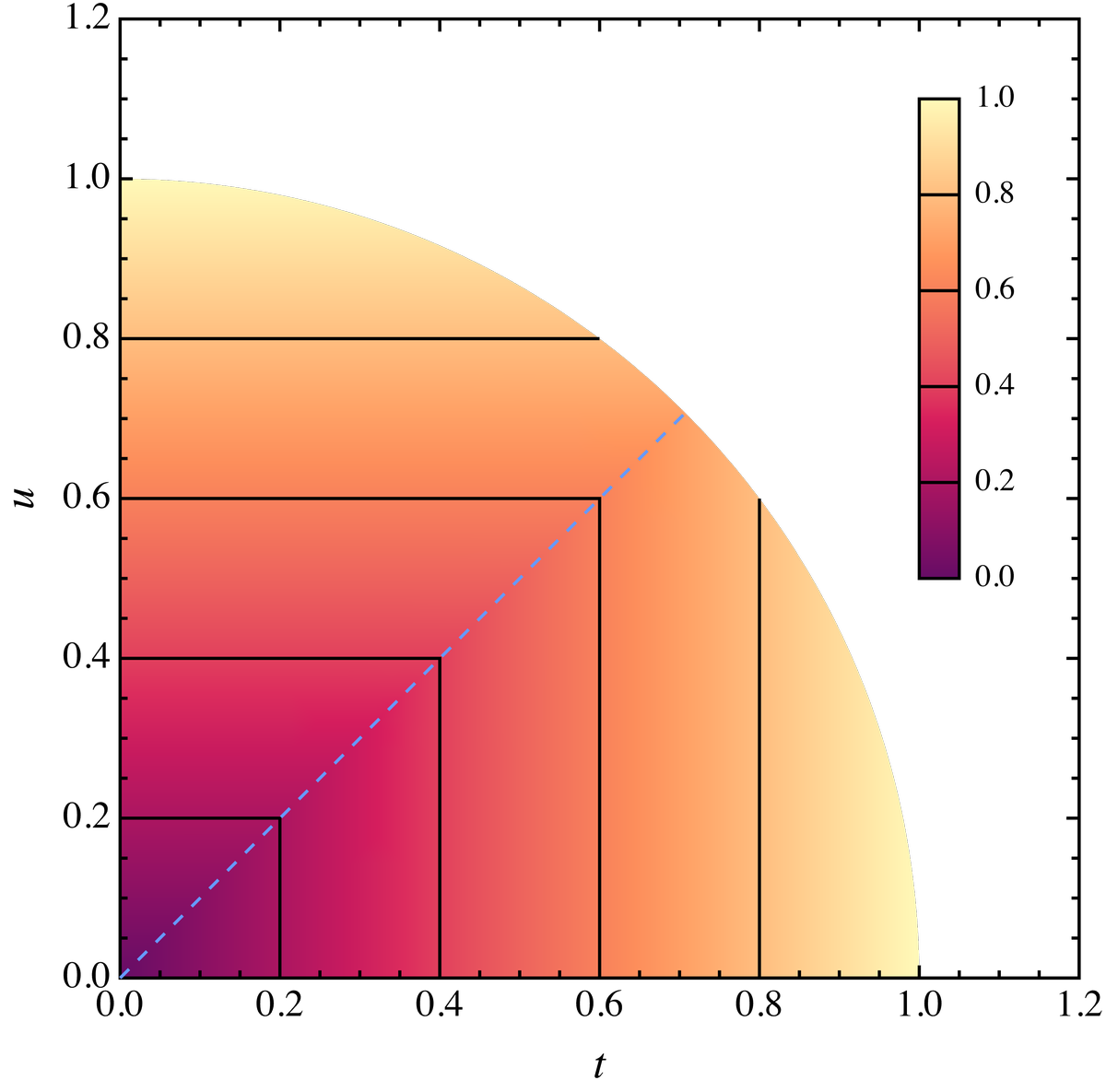}
    \caption{Output of the function $\max$ over the domain where the joint pdf $f(u,t)>0$.}
    \label{fig:maxvalue}
\end{figure}
The probability-density of a certain output $Z$ of the $\max$ function is given by the integral over the contour of $Z$ of the joint PDF. Since the top-left and the bottom-right piece of the graph are mirror symmetric (around the blue dashed line) and the joint PDF is also mirror symmetric around the same axis. Therefore, the function is split into two parts, where we only integrate in the bottom-right part and multiply the result by 2. The bottom-right part (with respect to the blue line) is described as:
\begin{align}
f_Z(w) &= \begin{cases}
\int \limits_0^w f(w,t) dt  & 0\leq w\leq \frac{1}{\sqrt{2}} \\
 \int \limits_0^{\sqrt{1-w^2}} f(w,t) dt  & \frac{1}{\sqrt{2}}\leq w\leq 1
 \end{cases} \\
f_Z(w) &= \begin{cases}
 \frac{4 \csc^{-1}\left(\sqrt{\frac{1}{w^2}-1}\right)}{\pi } & 0 \leq w \leq \frac{1}{\sqrt{2}} \\
 2 & \frac{1}{\sqrt{2}} \leq w \leq 1\\
\end{cases}
\end{align}
We can visualize the PDF of the output of the $\max$ function, which is shown in figure\ \ref{fig:maxoutput}.
\begin{figure}[htbp!]
    \centering
    \includegraphics[width=0.5\linewidth]{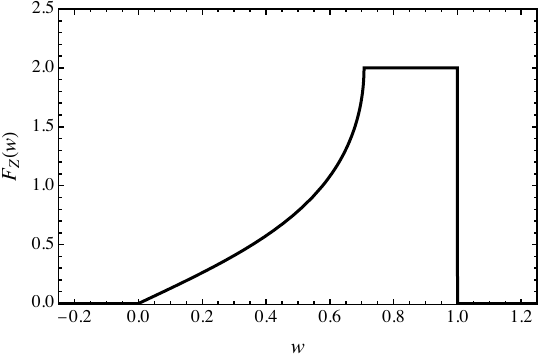}
    \caption{pdf of the max function ($f_Z(w)$).}
    \label{fig:maxoutput}
\end{figure}
Referring back to equation \eqref{eq:theta_uv}, the $\acos$ of the $\max$ function should be considered. Therefore, we transform the PDF of the $\max$ function with the $\Theta=\acos(w)$ function. The input domain is $0\leq w \leq 1$, and the output domain is $0\leq\theta\leq\pi/2$, in which the function monotonically decreases. In general we have:
\begin{align}
    F_\Theta(\theta) &= P(\Theta \leq \theta) = P(\acos(w) \leq \theta) \\
     \acos(w) \leq \theta &\Rightarrow w \geq \cos(\theta) \\
     F_\Theta(\theta) &= P(w \geq \cos(\theta)) = \int \limits_{\cos(\theta)}^{1} f_Z(w) dw
\end{align}
From the CDF ($F_\Theta(\theta)$), we differentiate to obtain the PDF ($f_\Theta(\theta)$), while using Leibniz' rule of integration:
\begin{align}
    f_\Theta(\theta) &= \frac d{d\theta} \int \limits_{\cos(\theta)}^{1} g(w) dw = -f_Z(\cos(\theta)) \frac{d\cos(\theta)}{d\theta} \\
    & = f_Z(\cos(\theta)) \sin(\theta)
\end{align}
The limits have to be transformed such that $0 \leq w \leq \frac{1}{\sqrt{2}}$ corresponds to $\pi/4 \leq \theta \leq \pi/2$ and $\frac{1}{\sqrt{2}} \leq w \leq 1$ corresponds to $0\leq \theta\leq\pi/4$. Using this and filling in $f_Z$ we find:
\begin{align}
    f_\Theta(\theta) &= \sin(\theta)\begin{cases}
 \frac{4 \csc^{-1}\left(\sqrt{\frac{1}{\cos(\theta)^2}-1}\right)}{\pi } & \pi/4 \leq \theta \leq \pi/2 \\
 2 & 0\leq \theta\leq\pi/4
\end{cases}
\end{align}
Employing trigonometric identities, the expression is simplified to:
\begin{align}
    f_\Theta(\theta) &= \begin{cases}
 2 \sin (\theta ) & 0\leq \theta \leq \frac{\pi }{4} \\
 \frac{4}{\pi} \asin(\cot (\theta )) \sin(\theta) &
   \frac{\pi }{4}\leq \theta \leq \frac{\pi }{2}
\end{cases}
\end{align}
The found distribution can now be compared with synthetically-generated data, which is shown in figure \ref{fig:oloiddist}.
\begin{figure}[htbp!]
    \centering
    \includegraphics[width=0.65\linewidth]{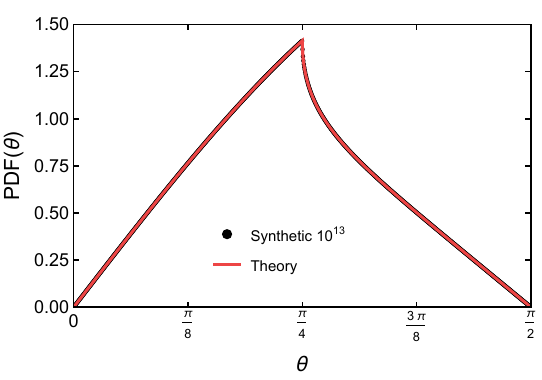}
    \caption{Probability density of the alignment of the pointing vector of a randomly-oriented oloid, taking into consideration the symmetries. Synthetic data comes from numerically simulating this process using equations \ref{eq:algo1}--\ref{eq:algo6} $10^{13}$ times to obtain a smooth pdf, which matches the theoretical derivation.}
    \label{fig:oloiddist}
\end{figure}

\end{appen}\clearpage

\bibliographystyle{jfm}
\bibliography{jfm}

\end{document}